\documentclass[journal]{IEEEtran}
\IEEEoverridecommandlockouts
\usepackage{cite}
\usepackage{amsmath,amssymb,amsfonts,amsthm}
\usepackage{algorithmic}
\usepackage{graphicx}
\usepackage{textcomp}
\usepackage{comment}
\usepackage{url}
\usepackage{authblk}
\usepackage{array}
\usepackage{multirow}
\usepackage{subfigure}
\usepackage[ruled,linesnumbered]{algorithm2e}

\usepackage[table]{xcolor}
\usepackage{booktabs}

\begin{document}

\title{\textit{CoRaiS}: Lightweight Real-Time Scheduler for Multi-Edge Cooperative Computing}  
\author{Yujiao Hu,~\IEEEmembership{Memeber,~IEEE,} Qingmin Jia, Jinchao Chen, Yuan Yao, Yan Pan, Renchao Xie,~\IEEEmembership{Senior Memeber,~IEEE,} F.Richard Yu,~\IEEEmembership{Fellow,~IEEE}
\thanks{Yujiao Hu, Qingmin Jia and Renchao Xie are with Future Network Research Center, Purple Mountain Laboratories, Nanjing 211111, China (email: huyujiao@pmlabs.com.cn, jiaqingmin@pmlabs.com.cn, renchao\_xie@bupt.edu.cn).

Jinchao Chen and Yuan Yao are with the School of Computer Science, Northwestern Polytechnical University, Xi'an 710029, China (email: cjc@nwpu.edu.cn, yaoyuan@nwpu.edu.cn).

Yan Pan is with Science and Technology on Information Systems Engineering Laboratory, National University of Defense Technology, China (email:panyan@nudt.edu.cn)

Renchao Xie is also with the State Key Laboratory of networking and Switching Technology, Beijing University of Posts and Telecommunications, Beijing 100876, China.

F. Richard Yu is with the Department of Systems and Computer Engineering, Carleton University, Ottawa, Canada (e-mail: richard.yu@carleton.ca).  

Corresponding authors: Qingmin Jia, Jinchao Chen

This work was supported in part by the National Natural Science Foundation of China (Grant Number: 92367104, 92267301), the Purple Mountain Talents-Jiangning Baijia Lake Plan Program (Grant Number: 74072203-3). 

Copyright (c) 20xx IEEE. Personal use of this material is permitted. However, permission to use this material for any other purposes must be obtained from the IEEE by sending a request to pubs-permissions@ieee.org.
}
}

\markboth{Journal of \LaTeX\ Class Files,~Vol.~14, No.~8, August~2021}%
{Shell \MakeLowercase{\textit{et al.}}: A Sample Article Using IEEEtran.cls for IEEE Journals}


\maketitle

\begin{abstract}
Multi-edge cooperative computing that combines constrained resources of multiple edges into a powerful resource pool has the potential to deliver great benefits, such as a tremendous computing power, improved response time, more diversified services. However, the mass heterogeneous resources composition and lack of scheduling strategies make the modeling and cooperating of multi-edge computing system particularly complicated. This paper first proposes a system-level state evaluation model to shield the complex hardware configurations and redefine the different service capabilities at heterogeneous edges. Secondly, an integer linear programming model is designed to cater for optimally dispatching the distributed arriving requests. Finally, a learning-based lightweight real-time scheduler, \textit{CoRaiS}, is proposed.  \textit{CoRaiS} embeds the real-time states of multi-edge system and requests information, and combines the embeddings with a policy network to schedule the requests, so that the response time of all requests can be minimized. Evaluation results verify that \textit{CoRaiS}  can make a high-quality scheduling decision in real time, and can be generalized to other multi-edge computing system, regardless of system scales. Characteristic validation also demonstrates that \textit{CoRaiS} successfully learns to balance loads, perceive real-time state and recognize heterogeneity while scheduling.


\end{abstract}


\section{Introduction}
\begin{figure}[t]
	\centering
	\includegraphics[width=0.8\columnwidth]{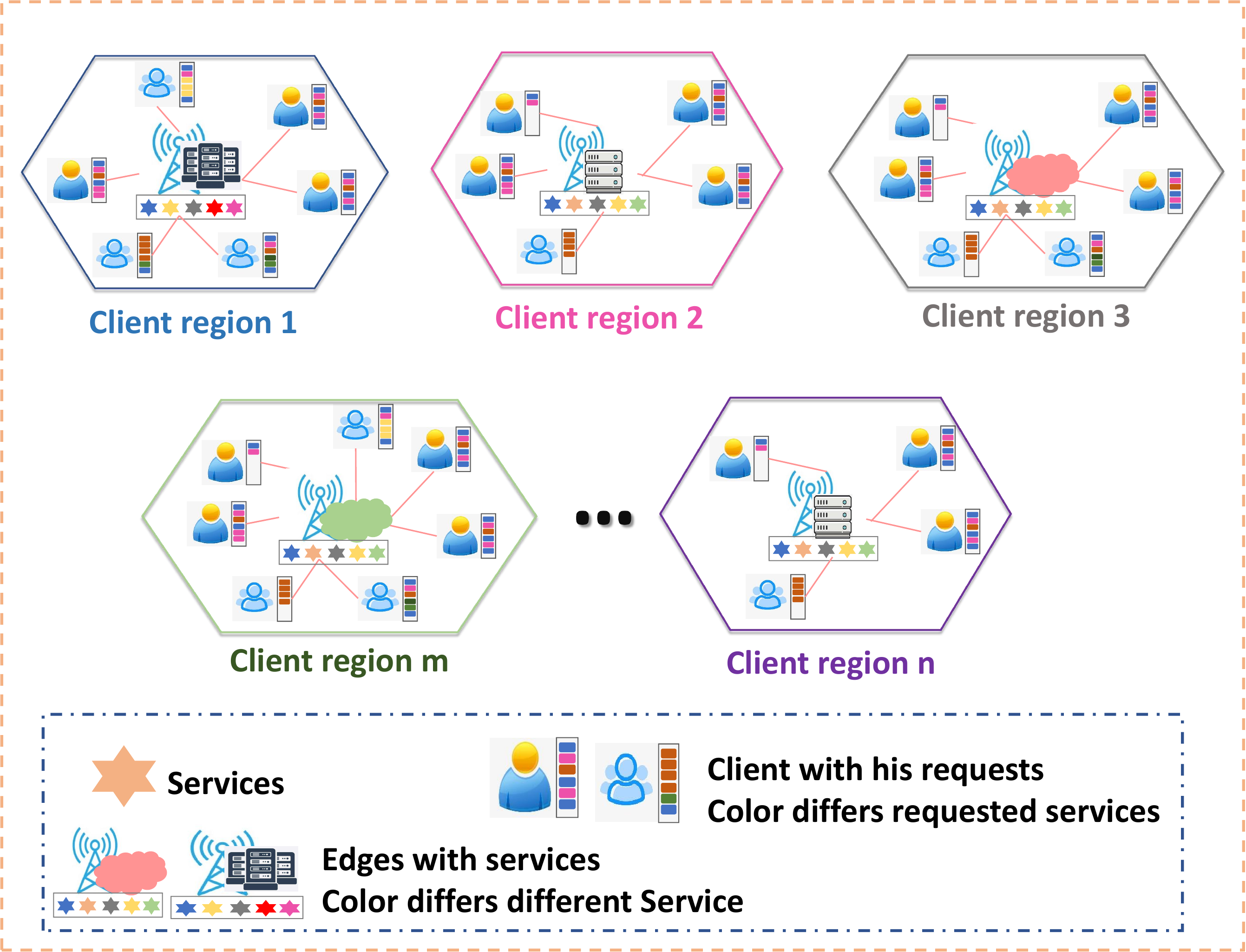} 
	\caption{The illustration of unbalanced workloads and resource utilization of edge computing.}
	\label{fig:edge-computing}
\end{figure}

Edge computing brings computation and storage resources closer to the sources of data, facilitating the processing of client data at the network periphery while meeting stringent response time requirements. Some great progresses have been achieved, especially in the mobile edge computing \cite{mao2017survey,mach2017mobile}. However, practical applications often reveal challenges. As shown in Fig.~\ref{fig:edge-computing}, each edge hosts a diverse set of services, and it typically serves multiple clients. The distribution of clients among edges exhibits non-uniformity, with variations in both the number of requests submitted by each client and the specific service they require. This complexity can potentially degrade service quality, especially when an edge is inundated with an excessive number of client requests.

The multi-edge cooperative computing combining constrained resources of multiple edges into a powerful resource pool can provide more diversified services and ensure sufficient computing and storage resources. Therefore, it has higher probability to meet the requirements by computation-intensive and latency-critical requests, and improve the average utilization of edges and response time of requests. The edge-cloud system can be regarded as a special case of multi-edge cooperative computing, because the cloud can be considered as an edge with significantly enhanced computational capabilities.

Many researchers are interested in the multi-edge cooperation system, and have proposed some algorithms \cite{10025811,farhadi2021service,han2021tailored,ren2022edgematrix,gao2018spotlight,mirhoseini2018hierarchical,chen2023scheduling} to schedule independent requests and requests with logical execution order in the system. They usually make some hypothesises to support their research. Firstly, only CPU is configured on the edges, and the number of calculations required to respond to each request is known. With such two conditions, the response time of each request can be obtained by dividing the calculation number by the CPU frequency. Secondly, the request arrival patterns follow known probability distributions, such as poisson distribution, multinomial distribution, etc. With such assumption, many statistical theories and models can be applied to the multi-edge scheduling problems. 

However, the assumptions contradict many practical situations.  \textit{(\romannumeral1) The heterogeneous edges are configured with various computing units, such as CPUs and GPUs.} Especially, with increasing requests related to deep learning accessing to edges, it has become a major trend to deploy GPUs at edges. 
\textit{(\romannumeral2) It is difficult to estimate the calculation number required by requests.} Because, when the corresponding processing code is black-box, the calculation number cannot be predicted in advance; when the code is white-box but includes some loop operation and judgment statements, the timing of jumping out of loops and the judgment results are also unpredictable, which makes the calculation numbers impossible to be estimated. 
\textit{(\romannumeral3) The arrival pattern of requests is almost unpredictable.} Each client has its own unique request generation pattern. Though the pattern of a single client can be estimated, it is difficult to analyze the composite pattern of multiple clients. 
Besides that, another two phenomenons should be focused on, i.e. during the execution process of a service, it will occupy multiple resources such as CPU, GPU, and Memory simultaneously, and the quality of service (QoS) varies across edges\footnote{The Quality of Service (QoS) for a specific service exhibits variations among heterogeneous edges due to differences in their hardware configurations. Additionally, even among homogeneous edges with identical hardware configurations, the QoS may differ because they allocate different resources to the service. For instance, if two edges share the same hardware setup, one may assign a 2-core CPU to the service while the other allocates a 4-core CPU. As a result, the QoS of the service will vary between these two edges.}. 

Based on the analysis above, we can identify that the multi-edge cooperative scheduling faces three challenges. 
\textit{(\romannumeral1) Multi-edge cooperative computing system state modeling:} The diverse hardware composition and different resource allocation schemes for services at edges pose a severe problem for modeling system state. The QoS of edges would not be fairly evaluated, unless the edges heterogeneity can be shielded and an unified QoS evaluation method can be built. Meanwhile, the system state that includes remaining resource of CPU/GPU/Memory at edges keeps changing as the arrival, running, and completion of requests. Dispatching requests to edges based on the perfect initial state, like traditional approaches, will make the scheduling solution deviate from the optimal at the current state. Therefore, it is challenging but necessary to build a system-level state model that supports edges unified modeling and perceives dynamic changes of resource states. 
\textit{(\romannumeral2) Multi-edge cooperative scheduling formulation:} Most previous theoretical models were designed based on probability distribution assumption of requests and single computing hardware assumption of edges \cite{ma2020cooperative,workload2021}, which may not accurately reflect real-world application scenarios. Some learning-based approaches \cite{han2021tailored,ren2022edgematrix} are also studied to optimize the scheduling behaviours on some specific datasets and specially constructed multi-edge network. However, the generalization ability\footnote{Here the generalization ability refers to the pre-trained scheduling models can help make effective decisions on unseen applications and multi-edge networks.} of the approaches is not good. Once the application and network environment change, a large amount of data (with expert knowledge sometimes) must be collected to retrain the scheduling model, while sometimes it is difficult to collect efficient data and expert knowledge. Therefore, a new mathematical formulation is required to reveal the essence of multi-edge cooperative scheduling problem and to guide scheduling algorithm research for requests with any arrival pattern. 
\textit{(\romannumeral3) Real-time scheduler designing:} The multi-edge scheduling problem for multiple requests is essentially a combinatorial optimization problem with some constraints. The search space of such problem is huge and will grow as the numbers of edges and requests increase. Computing an optimal solution in the huge search space is theoretically time-consuming. The previous works usually took a long time to make the scheduling decision. However, in practice, only methods that support real-time scheduling can ensure the efficient operation of the multi-edge cooperative computing system. 

To cope with the inherent challenges mentioned above, we first propose a system-level state evaluation model to express the service-oriented performance and workload of edges at any scheduling period. In this way, the important and differentiated performance characteristics closely related to edge scheduling are preserved, and the heterogeneous configuration of edges can be ignored when making scheduling decisions. Secondly, based on the system-level state evaluation model, we provide a new integer linear programming formulation for the multi-edge cooperation scheduling, which can be a good starting point to inspire solver searching or scheduling algorithms design. Finally, we propose \textit{CoRaiS}, a reinforcement learning based lightweight real-time scheduler for multi-edge cooperative computing system. Given a high-level goal to minimize the response time of all requests, \textit{CoRaiS} automatically learns a sophisticated system-level real-time scheduling policy. The policy can be directly generalized to other applications and multi-edge networks. 

The main contributions of this paper can be summarized as follows. 
\begin{itemize}
	\item A system-level state evaluation model is built to capture important features closely related to scheduling across edges, including service-oriented performance feature and workload feature.  
	\item The multi-edge scheduling problem is presented as a new integer linear programming formulation, which will support the designing and optimization of scheduling algorithms. 
	\item A lightweight attention-based scheduler (\textit{CoRaiS}) is proposed to minimize the response time over all requests distributed at edges. The scheduler can provide a high-quality near-optimal solution in real time, irrespective of request arrival patterns and system scales.
\end{itemize}

\textcolor{black}{
The following paper is structured as follows: Section~\ref{section:related-work} presents an overview of related work. In Section~\ref{section:use-cases}, we provide two use cases to illustrate the key aspects of request scheduling in a multi-edge cooperative computing system. Sections~\ref{section:system-model} and~\ref{section:outerformulation} delve into the designed model of the multi-edge cooperative computing system and the formulation of multi-edge scheduling, respectively. Following this, Section~\ref{section:CoRaiS} elaborates on the architecture of the proposed CoRaiS system. Simulation experiments and prototype-based experiments are detailed in Sections~\ref{section:simulation} and~\ref{section:prototype}, respectively. Finally, the conclusion in Section~\ref{section:conclusion} summarizes the findings and contributions of the paper.}

\section{Related Work}\label{section:related-work}
\textcolor{black}{
Previous studies have delved into the formulation and optimization of scheduling problems within multi-edge or edge-cloud cooperative computing systems, yielding significant contributions \cite{comprehensive2019,scheduling2018,task2019}. Typically, these works categorize dispatched requests into two groups: independent requests and requests with a logical execution order, subsequently designing specific scheduling algorithms for each category.}
\textcolor{black}{
\subsection{Independent Requests Scheduling}
Numerous efforts \cite{he2018s,farhadi2021service,han2021tailored,ren2022edgematrix,poularakis2019joint,ma2020cooperative,workload2021,tan2017online,han2019ondisc} have primarily focused on independent requests. Most of them formulate the request scheduling problem within multi-edge cooperative system as integer linear programming problem, and make some assumptions to enable the formulations.  For instance, \cite{he2018s,farhadi2021service} assume known computation requirements and average arrival rates, formulating the problem as an integer linear program and proving its NP-hardness. They further propose heuristic approximation algorithms based on linear program relaxation and rounding to address the issue. Other works like \cite{ma2020cooperative,workload2021} make assumptions about request arrival patterns and computation requirements to address scheduling challenges.  The works \cite{tan2017online,han2019ondisc} assume weightings indicating latency sensitivity and machine-dependent processing times for requests and propose online scheduling frameworks to minimize total weighted response time across all requests. The work \cite{poularakis2019joint} formulates a request routing problem under the assumption of computation capacity in multi-cell mobile edge computing networks, proposing a randomized rounding algorithm to minimize load to the cloud. }

\textcolor{black}{
Besides the heuristic methods, learning-based approaches have emerged as a significant contributor to addressing multi-edge cooperative computing challenges. For instance, 
the work \cite{han2021tailored} develops its dedicated experimental environment comprising physical servers and leased clouds, introducing a coordinated multi-agent actor-critic algorithm for decentralized request dispatch. This algorithm is evaluated using real-world workload traces from Alibaba. Similarly, \textit{EdgeMatrix} \cite{ren2022edgematrix} presents a learning-based scheduling model aiming to maximize throughput while ensuring various Service-Level-Agreement priorities. A networked multi-agent actor-critic algorithm is proposed to customize resource channels and enhance system stability. The work \cite{zhang2024lsia3cs} proposes a task scheduling framework across edge clouds, namely LsiA3CS, which employs A3C \cite{mnih2016asynchronous} and heuristic guidance to achieve distributed, asynchronous task scheduling for large-scale IIoT. Aiming at optimizing the average response time of task scheduling and the average energy consumption of the system, a multi-objective task scheduling model is designed, and a task scheduling policy optimization algorithm based on improved asynchronous advantage actor-critic (A3C) is proposed in \cite{cheng2024multi}.}

\textcolor{black}{
Additionally, the optimization objective of researches are also varying. The work \cite{yang2023skypilot} proposes a fine-grained two-sided market via an intercloud broker, named SkyPilot to allow users to view the cloud ecosystem and schedule requests in an economic approach. The work \cite{aburukba2021heuristic} uses mixed integer programming to formulate the scheduling problem, which aims to satisfy the maximum number of requests given their deadline requirements. Moreover, the work presents a heuristic approach using the genetic algorithm to generate the scheduling decision. In the work \cite{li2021software}, an optimal control theory (OCTS) based method is introduced. It aims to achieve global optimal results, ensuring computational and transmission latency requirements are met while load balancing various vehicle tasks. The work \cite{nguyen2019evolutionary} proposes an evolutionary genetic (GA)-based optimization algorithm, aiming to strike a balance between task execution time and processing cost. The work \cite{deng2019parallel} presents a Dynamic Parallel Computing Offload and Energy Management (DPCOEM) algorithm. It utilizes Lyapunov optimization technology to minimize task response time and achieve near-optimal performance. The work \cite{huang2019bilevel} formulates a bilevel optimization problem and proposes ant colony system and monotonic optimization method to minimize the energy consumption of all requests with deadline. The work \cite{chen2020resource} introduces a resource-constrained task scheduling profit optimization algorithm (RCTSPO). It focuses on profit maximization while enhancing task scheduling time, reliability, and load balancing.
}

\begin{table*}
    \caption{\textcolor{black}{Synthesis table of related works}}
	\renewcommand{\arraystretch}{1.2}
	\begin{center}
		\setlength{\tabcolsep}{3pt}
    \begin{tabular}{p{1.2cm}|p{1.9cm}|p{2cm}|p{3.3cm}|p{2cm}|p{3.5cm}|p{2cm}}
    \toprule
        Related works & Computing resources & Assumptions of request arrival pattern  & Assumptions of execution time for each request & Problem formulations & Optimization objectives & Scheduling algorithms \\ \midrule
        \cite{he2018s,farhadi2021service} & CPU & The average arrival rate at time slots is known & N/A & Mixed integer linear program (MILP)  & Maximize the number of requests served per slot & Greedy algorithm and linear program \\ \hline
        \cite{ma2020cooperative} & CPU & The arrival pattern at each edge node follows a Poisson process & Requests' CPU cycles   follow exponential distribution. The  computing time is estimated via FDC, i.e. the maximum frequency of edges divides required CPU cycles. & Mixed integer non-linear programming problem (MINonILP) & Minimize the service response time and outsourcing traffic to central clouds & Iterative optimization algorithm  \\ \hline
        \cite{workload2021} & CPU & The average arrival rate at time slots is known & Requests'  CPU cycles are randomly sampled with a finite expection.  The  computing time is estimated via FDC. & Queuing game & Minimize the expected cost of each individual edge server & Cooperative queueing game approach \\ \hline
        \cite{tan2017online,han2019ondisc}  & CPU & N/A & Requests'  CPU cycles are known and an edge can execute at most one request at a time & Dual integer programming & Minimize the total weighted response time of all requests & Online algorithm with dispatching policy and the scheduling policy \\ \hline
        \cite{aburukba2021heuristic} & CPU & N/A & Requests'  CPU cycles are known, and the computing time is estimated via FDC. & MILP & Maximize the number of requests with deadlines & Genetic algorithm \\ \hline
        \cite{deng2019parallel} & CPU & The arriving probability at each slot is known & Requests'  CPU cycles are known, and the computing time is estimated via FDC. & MINonILP & Minimize response time and packet losses of tasks under the limitation of energy queue stability & Lyapunov-based dynamic parallel computing offloading \\ \hline
        \cite{nguyen2019evolutionary} & CPU & N/A & Requests'  CPU cycles are known, and the computing time is estimated via FDC. & MILP & Achieve a trade-off between execution time and monetary cost to complete requests & Particle swarm optimization \\ \hline
        \cite{huang2019bilevel} & CPU & N/A & Requests'  CPU cycles are known, and the computing time is estimated via FDC. & MILP & Minimize the total energy consumption of all mobile users under the delay constraint & Ant colony and monotonic optimization \\ \hline
        \cite{chen2020resource} & CPU & N/A & Requests'  CPU cycles are known, and the computing time is estimated via FDC. & Petri Net & Load balancing and profit optimization & Heuristic algorithm \\ \hline
        \cite{han2021tailored} & CPU & Specified based on dataset & Obtained from dataset & Markov game formulation  & Balance the workloads among edge and offload some requests to the cloud & Coordinated multi-agent actor-critic \\ \hline
        \cite{ren2022edgematrix} & CPU & Specified based on dataset & Obtained from dataset & Markov game formulation and MILP & Maximize the throughput while guaranteeing various SLA priorities & Multi-agent actor-critic algorithm  \\ \hline
        \cite{zhang2024lsia3cs} & CPU & Specified based on dataset & Obtained from dataset & Markov game formulation & Maximize the long-term throughput rate & Actor-critic algorithm \\ \hline
        \cite{mao2019learning} & CPU & Stochastic requests arrivals & The average execution time is known (from dataset) & Markov decision process (MDP) & Minimize average request completion time & Policy gradient \\ \hline
        \cite{ni2020generalizable} & CPU & N/A & Requests'  CPU cycles are known, and the computing time is estimated via FDC. & Search problem & Maximize system throughput & Graph-aware encoder-edecoder reinforcement learning model \\ \hline
        \cite{zhang2023multi} & CPU & N/A & Requests'  CPU cycles are known, and the computing time is estimated via FDC. & MDP & Minimize the long-term average delay and energy consumption, and maximize the system throughput  & Value decomposition multi-agent deep Q learning \\ \hline
        \cite{mirhoseini2017device,gao2018spotlight,mirhoseini2018hierarchical} & GPU and CPU & N/A & Measured in the hardware environment & MDP & Minimize the training time of a deep neural network & Reinforcement learning (RL) \\ \hline
        \cite{10246420} & GPU and CPU & N/A & Measured in the hardware environment & MILP and MDP & Minimize the training time of deep neural networks & Multi-agent RL \\ \midrule
        Ours & Service-oriented any resource combinations & N/A & Estimated through the proposed system-level state evaluation model & MILP & Minimize the average response time of computing requests & RL \\ \bottomrule
    \end{tabular}
    \end{center}
    \label{table:synthesis-table}
\end{table*}

\textcolor{black}{\subsection{Scheduling Requests with Logical Execution Order}
Dispatching requests with logical execution orders constitutes an important area of study within multi-edge cooperative computing. The dependencies among such requests are often represented as directed acyclic graphs (DAGs), leading to what is commonly referred to as the DAG scheduling problem. }

\textcolor{black}{
The work \cite{mao2019learning} introduces \textit{Decima}, which devises novel representations for requests' dependency graphs. It employs scalable reinforcement learning models to learn workload-specific scheduling algorithms autonomously, aiming to minimize average request completion time. In \cite{ni2020generalizable}, computing devices are allocated to continuous data flows within a large distributed system. The paper presents a graph-aware encode-decoder framework designed to learn a generalizable resource allocation strategy. The work \cite{zhang2023multi} proposes an online concurrent user request scheduling mechanism based on multi-agent deep reinforcement learning to optimize the long-term average delay and energy consumption.}

\textcolor{black}{
In addition to traditional request scheduling, the optimization of device placement for decomposed deep neural networks (DNNs) has also garnered attentions in the multi-edge cooperative scheduling domain, inspired by advancements in neural network-based intelligent large models. Studies also treat the architecture of a DNN model as a DAG, with device placement specifying how each operation in the model should be matched to networked heterogeneous CPU and GPU devices. For example, \textit{Mirhoseini et al.} \cite{mirhoseini2017device} propose a sequence-to-sequence model that optimizes device placement for TensorFlow computational graphs. \textit{Spotlight} \cite{gao2018spotlight} is devised to determine optimal device placement for DNN training. It formulates the problem as a Markov decision process with multiple stages and employs a novel reinforcement learning algorithm based on proximal policy optimization.
A two-level hierarchical model \cite{mirhoseini2018hierarchical} is also introduced for device placement in neural networks containing tens of thousands of operations. 
The work \cite{10246420} proposes TapFinger, a distributed scheduler for edge clusters that minimizes the total completion time of ML tasks through co-optimizing task placement and fine-grained multi-resource allocation. }

\subsection{Related Work Analysis}
\textcolor{black}{We analyze the related works from perspectives of computing resources, assumptions on request arrival patterns and computation requirements, problem formulations, optimization objectives and scheduling algorithms, and summarized them in Table~\ref{table:synthesis-table}. }
\textcolor{black}{Delving into the details of previous researches, we can find that previous research often made assumptions based on simplified scenarios, such as only having CPUs on the edges and knowing the number of calculations required for each request. However, in reality, heterogeneous edges equipped with various computing units, including GPUs and TPUs, are becoming increasingly common. Moreover, with the rise in requests related to deep learning, deploying GPUs at the edges has emerged as a significant trend. This shift introduces a new layer of complexity. The interaction of multiple computing resources (CPU, GPU, TPU, etc.), each with its own capabilities and characteristics, coupled with the intricate execution logic of intelligent applications, poses challenges in accurately estimating the running time of requests. 
Additionally, many works in the field make assumptions about request arrival patterns following known probability distributions like Poisson or multinomial distributions. However, the reality is that the arrival pattern of requests is often unpredictable. Each client may indeed have its own unique request generation pattern, making it challenging to analyze the composite pattern of multiple clients.} 

\textcolor{black}{
In contrast to existing works, our research liberates from many assumptions. Firstly, we extend computing devices to encompass a combination of CPUs and GPUs, facilitating the development of complex intelligent applications. We achieve this by constructing a novel system-level state evaluation model capable of evaluating the computation time of each request. This model is compatible with heterogeneous hardware configurations and diverse intelligent applications.
Secondly, we do not constrain the request arrival pattern. Instead, we design a new multi-edge cooperative framework and propose a corresponding lightweight scheduling algorithm to enable real-time scheduling of requests. Both the framework and algorithm are designed to accommodate any arrival pattern of requests.
}

\section{Motivating Use Cases}\label{section:use-cases}
\begin{figure}[t]
	\centering
	\includegraphics[width=0.8\columnwidth]{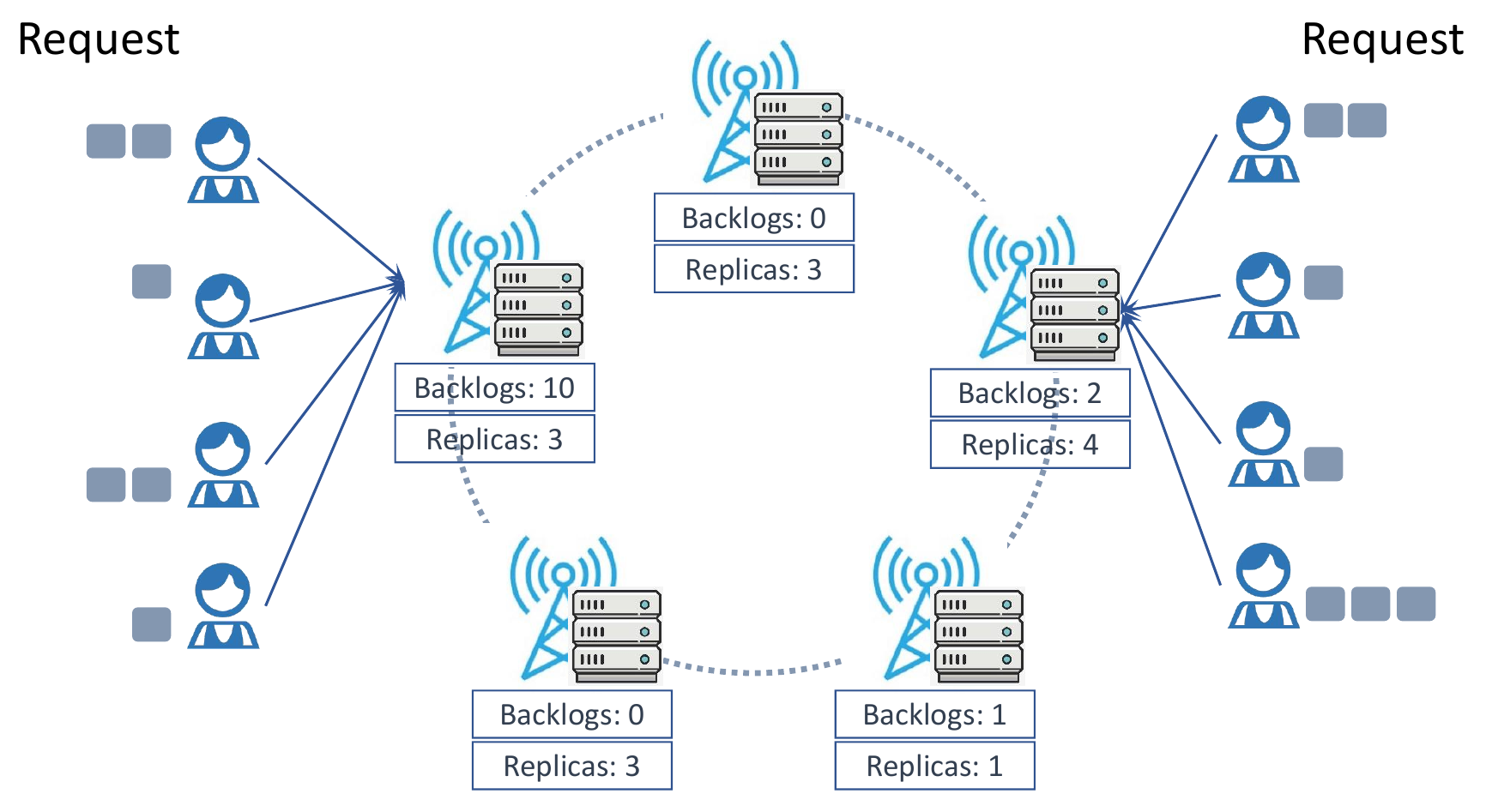} 
	\caption{\textcolor{black}{An illustration of the motivating use case. }}
	\label{fig:use-case-1}
\end{figure}

\begin{figure}[t]
	\centering
	\includegraphics[width=0.8\columnwidth]{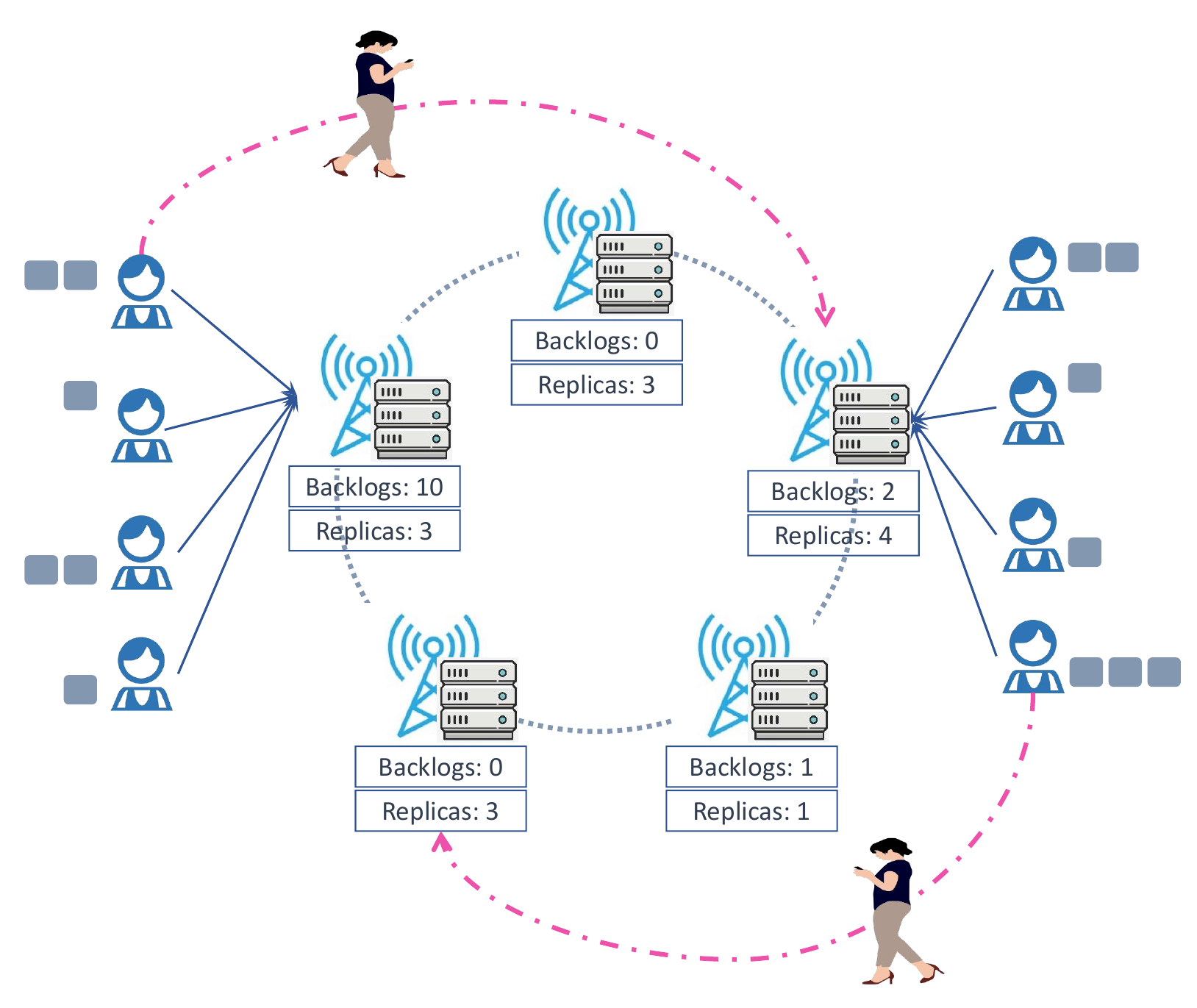} 
	\caption{\textcolor{black}{An illustration of the motivating use case with mobility. }}
	\label{fig:use-case-mobility}
\end{figure}

\textcolor{black}{
We present two use cases to illustrate the key points of request scheduling in multi-edge cooperative computing system. As shown in Fig.~\ref{fig:use-case-1}, five edges coordinate to serve clients, but the service capability and system status of edges are different, i.e., the hardware and software configure may be heterogeneous, and the backlogs and deployed service replicas can be different as well. Meanwhile, clients have different number of requests that need to be processed. Fig.~\ref{fig:use-case-mobility} considers client mobility into the use case, i.e. clients can move to another client region after submitting their requests. Request scheduling has to dispatch each request to an edge and make the total response time of requests minimized with consideration of the edge heterogeneity and client mobility. }

\textcolor{black}{
Multi-edge cooperative computing can also offer significant benefits in the context of content delivery networks (CDNs) \cite{stocker2017growing,zolfaghari2020content,jiang2021survey}, which are distributed networks of servers strategically positioned worldwide to deliver web pages, images, videos, and digital assets with high performance and availability. 
For instance, in video streaming applications, traditional CDN setups involve caching video content at edge servers for faster delivery. However, video content often requires transcoding into different formats or bitrates to suit various devices and network conditions. Multi-edge computing enables real-time video transcoding at edge servers. When a user requests a video, the edge server dynamically transcodes it into the appropriate format or bitrate based on the user's device and network conditions. This eliminates the demands for centralized transcoding servers and reduces latency by delivering optimized video directly from the edge server.
Similarly, in online gaming, CDNs optimize content delivery by caching game assets, updates, and patches at edge servers. Multi-edge computing can enhance this by distributing these assets across multiple edge locations, ensuring faster downloads and updates for players. Moreover, edge servers can dynamically adjust content delivery based on players' location, network conditions, and device capabilities, ensuring optimal performance for each player. 
}

\textcolor{black}{
However, there are two challenges when realizing the real-time effective scheduling. Firstly, 
there is a lack of effective business models or interoperability models to facilitate the cooperative interaction among edges and the precise execution of the scheduling decision. 
Because the traditional cloud business pattern, which submits all requests to the cloud management platform and lets the platform schedule the requests to specific servers, suffers from the huge data traffic consumption, the platform bottleneck and longer response time. Collaboration based on distributed consensus can lead to long-term negotiations at the edges, resulting in delayed scheduling decisions. 
Secondly, the scheduling space is $N^M$, where $N$ and $M$ refer to the number of edges and requests, respectively, indicating that the solution space will increase exponentially with the growth of edge numbers and request numbers. The huge scheduling space makes it difficult to search the optimized solution in time, while the practical scenario requires immediate solutions to enable the multi-edge system operation. }

\textcolor{black}{In the following, we will describe our interoperability models for the multi-edge cooperative computing system and formulate the scheduling problems in detail. }

\section{Multi-edge Cooperative Computing System Model}\label{section:system-model}

\subsection{Service-oriented Multi-edge Computing System Modeling}
\begin{figure}[t]
	\centering
	\includegraphics[width=0.8\columnwidth]{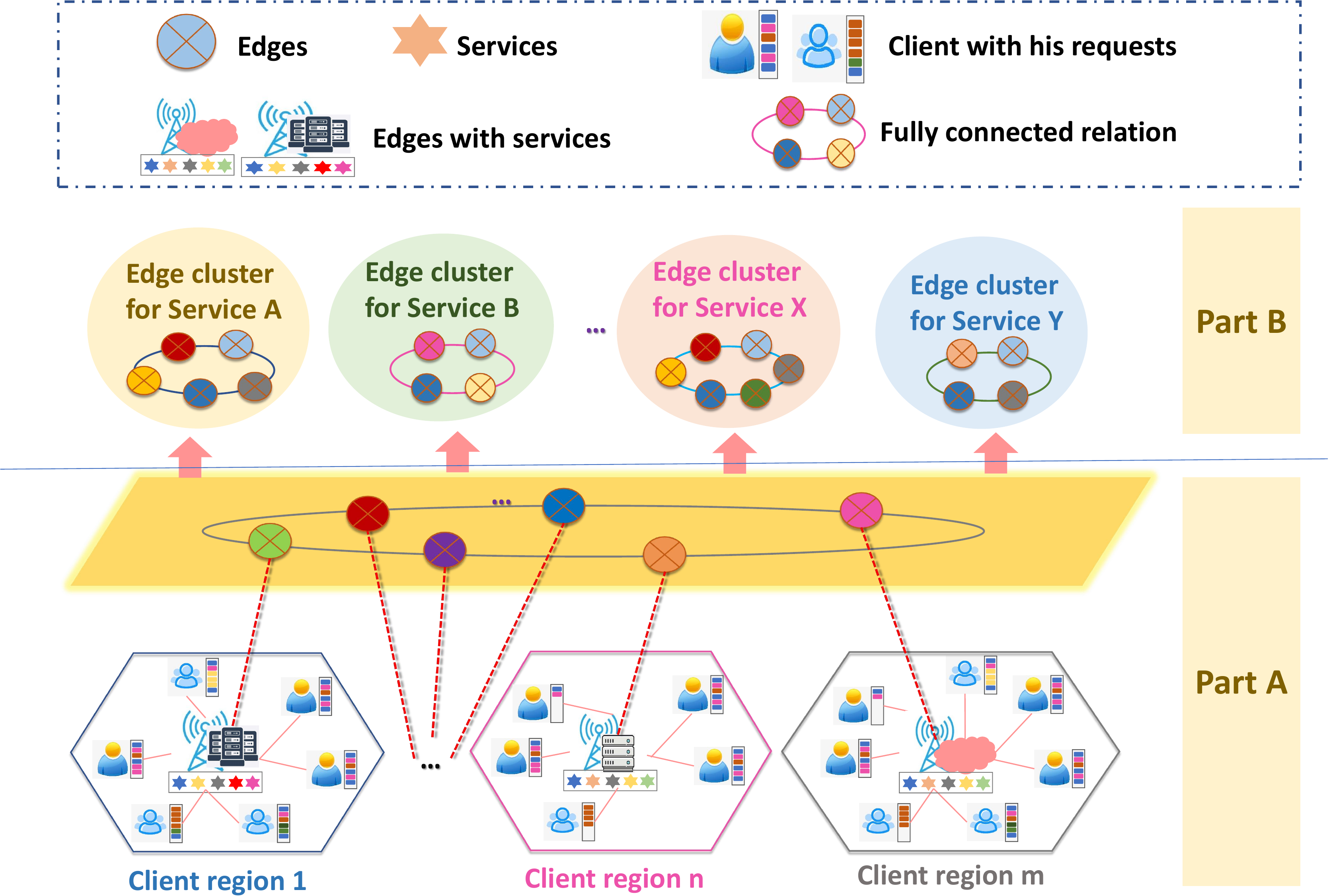} 
	\caption{Multi-edge cooperative computing system modeling. 
 \textit{Part A:} The edges are physically interconnected. Each edge has its dedicated client region and the services deployed on each edge operate independently. 
 \textit{Part B:} The multi-edge cooperative computing system is decomposed into multiple independent service-oriented subsystems, where the edges with same service are grouped.}
	\label{fig:mutli-edge-cooperative-computing-system}
\end{figure}

\textcolor{black}{
The multi-edge cooperative computing system consists of a set of network edges $\mathcal{E} = \{e_n\}_{n=1}^N$, and can provide diverse services $\mathcal{S} = \{s_k\}_{k=1}^K$, as shown in Fig.~\ref{fig:mutli-edge-cooperative-computing-system}. To enable service-oriented effective scheduling, we decompose the multi-edge cooperative system into multiple service-oriented subsystems, i.e.  $\{SR_k\}_{k=1}^K$,  where $SR_k=\{e_{k1}, e_{k2}, ..., e_{kn'}\}$, $e_{ki}$ denotes the $i^{th}$ edge with service $s_k$. }
With the decomposition, our focus can then shift towards addressing scheduling problem within each individual subsystem, since the process of scheduling over the whole multi-edge system can be decomposed into scheduling on each $SR_k$. For the sake of clarity, in the following discussions, we omit explicitly declaring the $s_k$-oriented. However, it's important to understand that all the described operations can be applied into any service-oriented subsystems.

\begin{figure}[t]
	\centering
	\includegraphics[width=0.9\columnwidth]{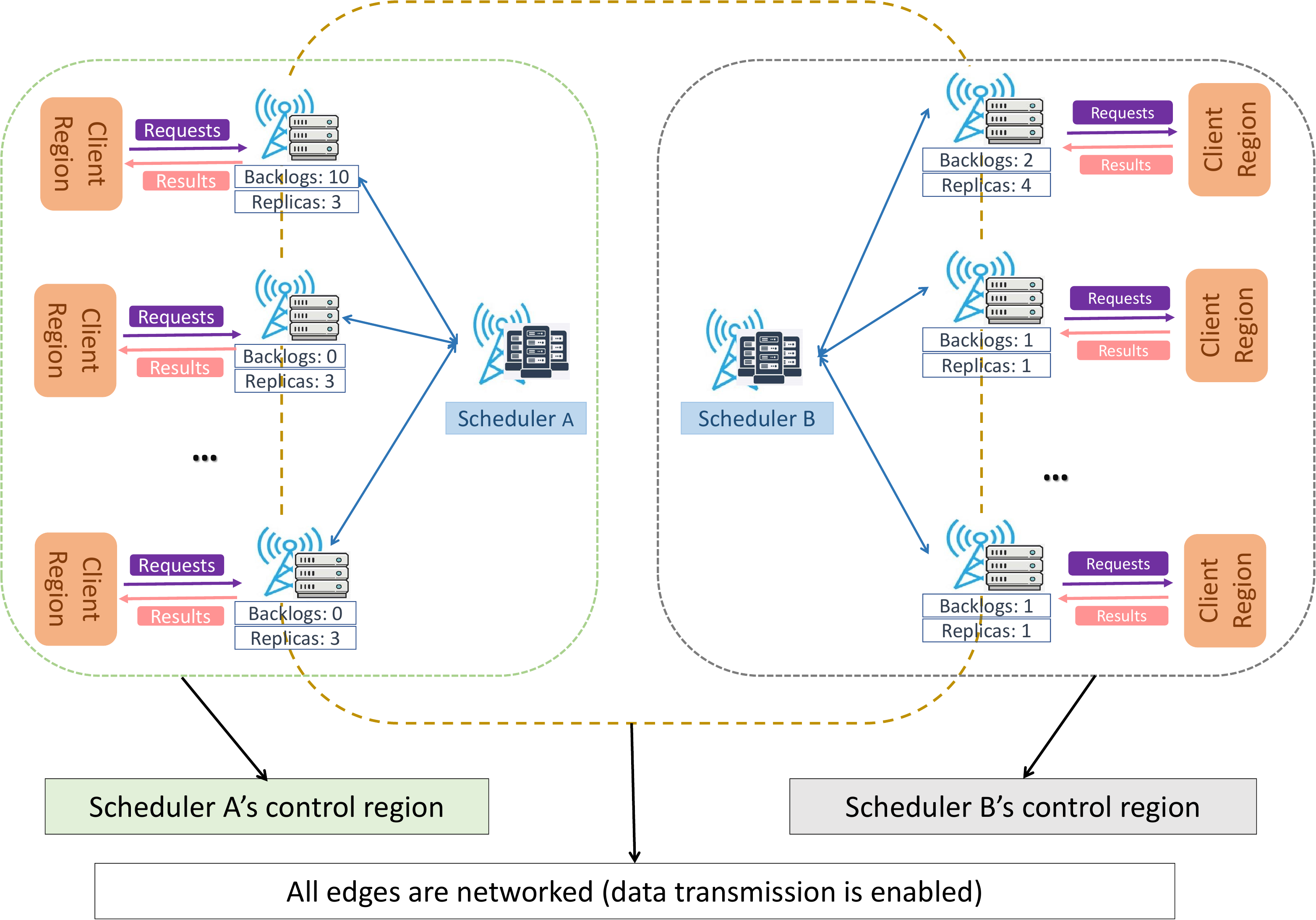} 
	\caption{\textcolor{black}{Territorial schedulers are introduced into the service-oriented multi-edge cooperative computing system to facilitate effective and real-time scheduling. }}
	\label{fig:territorial-scheduler-introduction}
\end{figure}

\textcolor{black}{
We further introduce territorial schedulers into the service-oriented multi-edge cooperative computing system. The novel multi-edge cooperative architecture is presented in Fig.~\ref{fig:territorial-scheduler-introduction}. All edges are networked, which enabling data transmission between any two edges. Each edge is under the exclusive management of a single scheduler, which decides the execution location of each received request in its control region.  Moreover, every edge within the system willingly collaborates with others and entrusts its scheduler's decision-making efficacy. Consequently, each edge actively processes all requests dispatched by the scheduler. This collective commitment ensures a seamless and efficient process, contributing to the overarching goal of achieving timely completion of all tasks.}

\textcolor{black}{
In addition, when an edge device seeks to join the system and contribute its computational resources, it must actively establish a connection with a base station. Subsequently, it is required to transmit registration messages to the territorial scheduler. These registration messages should include details such as deployed services, the number of corresponding replicas, computational capabilities, and geographical locations. 
Following the registration process, the edge device is expected to periodically send its status information to its designated scheduler, indicating its availability to handle incoming requests. If the information fails to be received continuously for a predetermined threshold number of times, the scheduler will infer that the edge device has gone offline, then if the edge wants to join the system again, it has to register like a new edge. 
By performing these operations, clients have the capability to submit requests to the edge device through either wireless or wired communication channels. Simultaneously, the scheduler is empowered to efficiently dispatch requests to the appropriate edge devices
}

\subsection{Interoperability Model of Multi-edge Cooperative Scheduling}
\subsubsection{\textcolor{black}{Concept definition}}~\textcolor{black}{Some concepts used in the interoperability model of multi-edge cooperative scheduling are described as follows.} 

\textcolor{black}{ \textit{\textbf{Request: }} A request typically comprises description text and physical input data. The description text outlines essential details such as the required service, client ID, source edge location from which the request originated, and the predicted edge location where the client will be located upon request completion. 
Input data refers to the actual data intended for processing. For instance, in an image classification request, the description text would specify the required classifier, while the related images would be uploaded as input data}

\textcolor{black}{
\textit{\textbf{Request brief: }} 
A request brief is essentially a data package containing only descriptive information about the request, devoid of detailed input data. For instance, in the context of an image classification request, the brief would encompass details such as the image sizes and the required classifier, while omitting the actual image content. Additionally, the brief includes information about the source edge location and the predicted edge location. 
The introduction of request briefs leads to a significant reduction in the size of data packages used for scheduling decisions. This reduction in package size has a direct impact on communication delay among edges and the scheduler, resulting in improved efficiency. 
}

\textcolor{black}{
\textit{\textbf{Scheduling decision: }}
It contains the information which edge will execute the requests. The edges are required to adhere to this decision, either responding to the requests locally or transferring them to other edges as directed.   
}

\subsubsection{\textcolor{black}{Interoperability scheduling model}}~\textcolor{black}{Based on the architecture shown in Fig.~\ref{fig:territorial-scheduler-introduction}, we further propose an interoperability scheduling model to enable the effective multi-edge cooperation. }

\textcolor{black}{The multi-edge cooperative scheduling processes involve eight steps:}
\textcolor{black}{
\textit{(\romannumeral1)} Clients submit requests and related data to edges, along with their predicted edge location after request completion.}
\textcolor{black}{
\textit{(\romannumeral2)} Edges receive requests and generate a request brief for each.}
\textcolor{black}{
\textit{(\romannumeral3)} Edges provide their schedulers with current service capacity information and the request briefs.}
\textcolor{black}{
\textit{(\romannumeral4)} The scheduler makes scheduling decisions based on edge status, request briefs, and designated scheduling algorithms.}
\textcolor{black}{
\textit{(\romannumeral5)} The scheduler informs edges of the decision, specifying the execution edge for each request.}
\textcolor{black}{
\textit{(\romannumeral6)} Based on the decision, edges locally handle the requests or transfer them to other edges along with the relevant data.}
\textcolor{black}{
\textit{(\romannumeral7)} The execution edges provide computing results feedback to the predicted edges.}
\textcolor{black}{
\textit{(\romannumeral8)} The predicted edges transmit the computing results to clients.}

\begin{figure}[t]
	\centering
	\includegraphics[width=0.9\columnwidth]{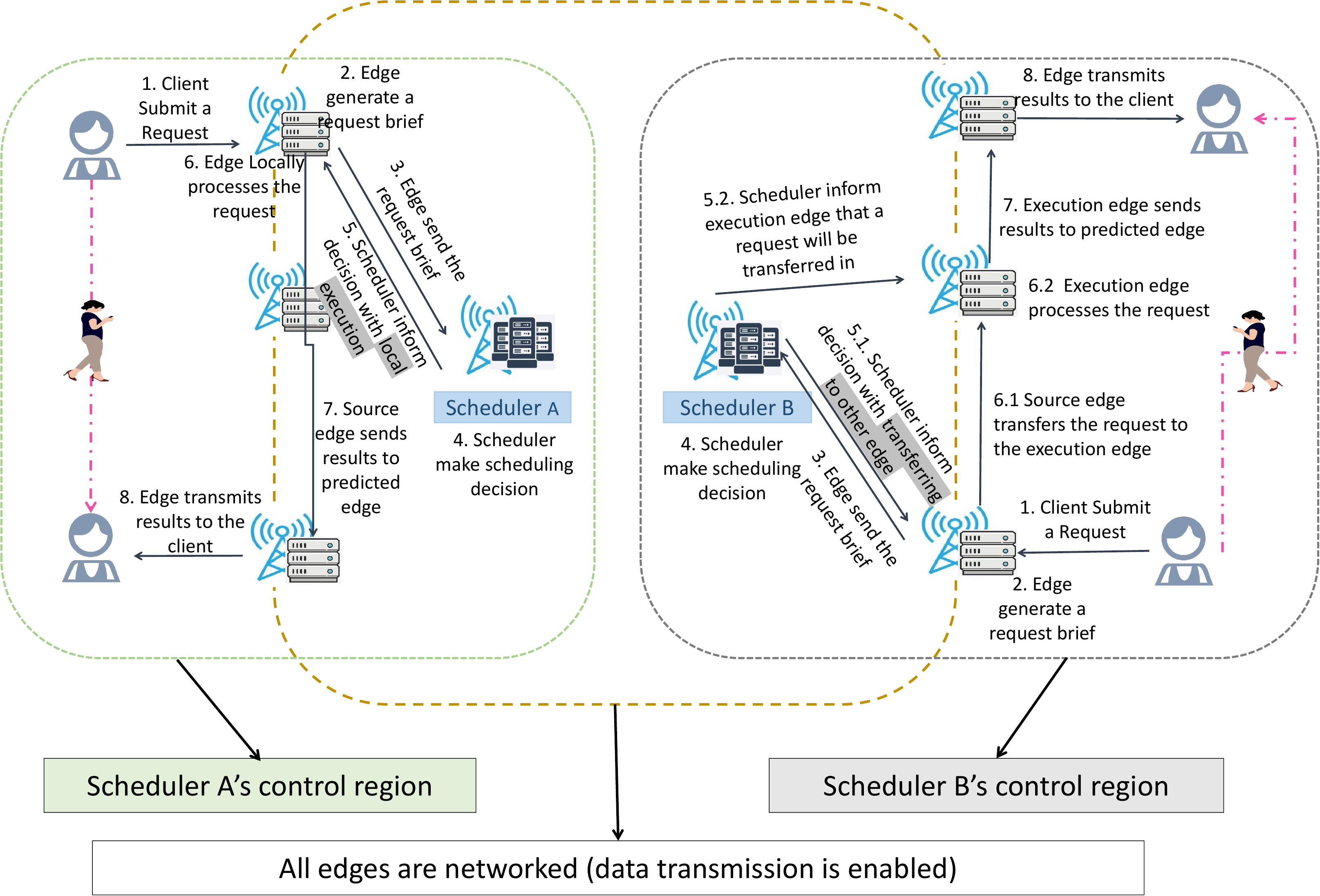}
	\caption{\textcolor{black}{An example of multi-edge cooperative scheduling considering client mobility within a single control region. In Scheduler A's control region, the submitted request is executed locally, indicating the source edge and the execution edge are the same. In Scheduler B's control region, the request is transferred to another edge for processing. In this scenario, Scheduler B must simultaneously inform both the source edge and the execution edge (actions 5.1 and 5.2). Subsequently, the source edge transmits all relevant data of the request to the execution edge (action 6.1), which then follows the scheduling decision and handles the request accordingly (action 6.2). Once the request is completed, the execution edge transmits the results to the prediction edge (action 7), which in turn provides feedback to the client (action 8).}
 } 
	\label{fig:Scheduling-process1}
\end{figure}

\begin{figure}[t]
	\centering
	\includegraphics[width=0.9\columnwidth]{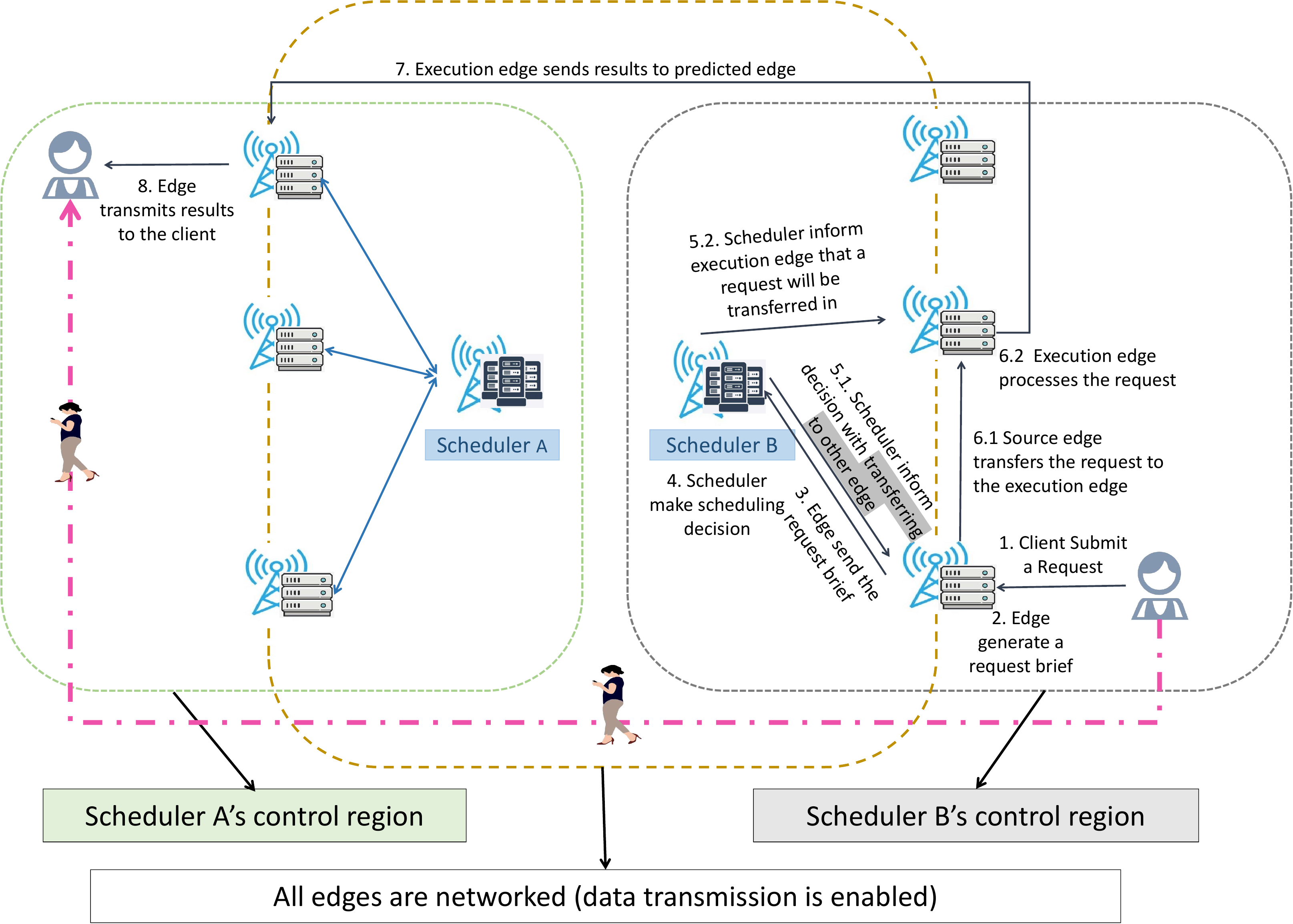} 
	\caption{\textcolor{black}{In a scenario involving multi-edge cooperative scheduling and client mobility across multiple control regions, schedulers are limited to scheduling edges within their respective control regions. When a request extends beyond this boundary, the scheduler dispatches it to one of its managed edges. Upon completion of the request, the execution edge transmits the results to the predicted edge via cross-region communications (action 7).} }
	\label{fig:Scheduling-process2}
\end{figure}

\textcolor{black}{
To provide a clearer illustration of the interaction processes between schedulers and edges supporting effective request response, we present two examples shown in Fig.~\ref{fig:Scheduling-process1} and Fig.~\ref{fig:Scheduling-process2}. These examples encompass scenarios involving client mobility both within the control region of a scheduler and beyond its boundaries.}

\section{Multi-edge Scheduling Formulation}\label{section:outerformulation}
In this section, we will formulate the multi-edge scheduling problem in detail. 

\subsection{System-level State Evaluation Model} \label{section:system-level-state-evaluation}
\begin{figure}[t]
	\centering
	\includegraphics[width=0.8\columnwidth]{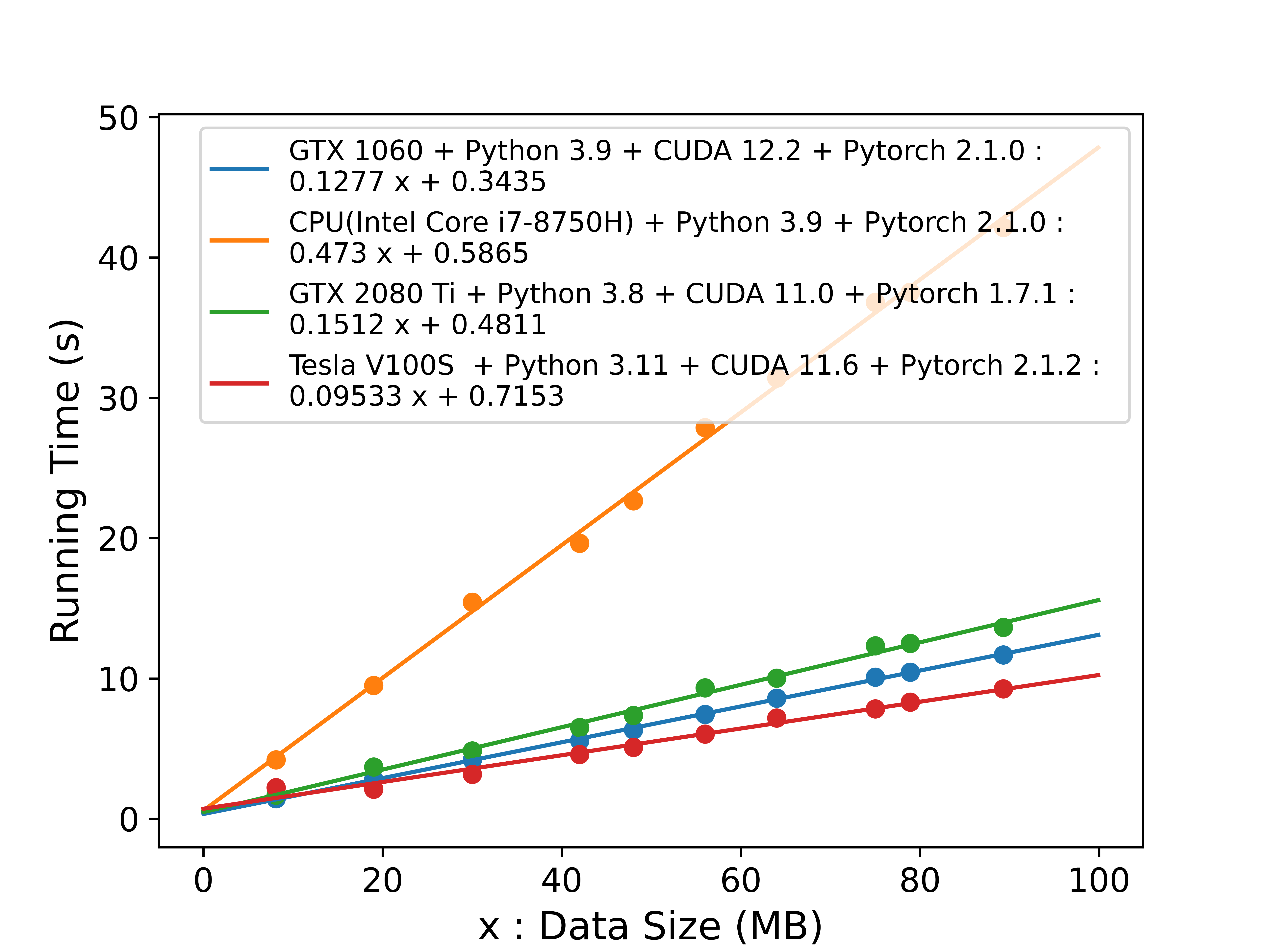} 
	\caption{\textcolor{black}{The linear relationship between running time and the size of data packets for \textit{Model Soups} \cite{pmlr-v162-wortsman22a} on heterogeneous devices. Points are actual statistical data, while lines are function relationships fitted based on the data. }} 
	\label{fig:soups}
\end{figure}
The system-level state evaluation model is inspired by some observations. 
The first key observation is that there are many popular artificial intelligence services, such as image classification and object detection, exhibiting a functional relationship between their response time and the size of the data packets to be processed. We select the most advance image classification model \textit{Model Soups} \cite{pmlr-v162-wortsman22a} as an example to illustrate the observation, and visualize the relationship of \textit{Model Soups} in Fig.~\ref{fig:soups}. In Fig.~\ref{fig:soups}, we can observe that the running time and data size of \textit{Model Soups} is linearly correlated, and the coefficients of the linear function are related to the configuration of computing devices. The most popular object detection model, \textit{YOLOv6} \cite{li2023yolov6,li2022yolov6} also cares about the computation efficiency. They reports the linear relationship between running time and data size for different vision of \textit{YOLOv6} when inferring on Tesla T4. 
Please note, in this paper, we refer to services that have a functional relationship between runtime and data size as \textit{ideal services}, and our study primarily focuses on this category. 
The second observation is that many advanced technologies, such as \textit{Docker} and \textit{Kubernetes}, empower one service to create multiple independent replicas on a single device and enable the service to specify the required resources. This reservation mechanism ensures that the resources allocated to each service replica are safeguarded against preemption by other processes, thereby stabilizing the QoS of the service.

Based on these observations, we build our system-level state evaluation model. The model consists of two parts: \textit{service-oriented performance estimation} and \textit{service-oriented workload evaluation}. 

\subsubsection{Service-oriented Performance Estimation} \label{section:performance-estimation}
Two indicators are used to consistently evaluate the service-oriented performance of edges, that are \textit{computation time estimation function} ($\phi(x)$) and \textit{service replica numbers} ($\zeta$). 
$\phi_i(x)$ is a function that depicts the relationship between the response time and the size of data packets to be processed at edge $e_i$, with $x$ denotes the size of the data packets. $\phi_i(x)$ can be approximated by fitting the relationship between data volume and the actual processing time of historical requests, like the fitting operation in Fig.~\ref{fig:soups}. There are some tools that can help establish the relationship, such as \textit{numpy.polyfit} and \textit{scipy.optimize.curve\_fit}. 
The hardware configuration also has a significant impact on the execution efficiency, causing $\phi(x)$ to vary across different edges. Fig.~\ref{fig:soups} illustrates the impact as well. Therefore, when establishing the relationship, only local historical data can be selected. 
$\zeta_i$ refers to the replica number of the service on edge $e_i$, which is a predefined system-level parameter to support service parallels. A larger $\zeta_i$ indicates edge $e_i$ can deal with more requests in parallel. 

By defining $\phi(x)$ and $\zeta$, we can focus on considering the primary performance factors of edges while formulating multi-edge cooperative scheduling problem and designing algorithms, and ignore the secondary factors (the heterogeneous hardware configuration and various resource allocation mechanism to services at edges). Furthermore, with $\phi_i(x)$, we can predict response times on any edges by inputting the request's data size. This eliminates the need to analyze black box or white box code to obtain the necessary computation numbers for the request and restricts the computing device to the CPU.

\subsubsection{Service-oriented Workload Evaluation} \label{section:workload-evaluation}
\begin{figure}[t]
	\centering
	\includegraphics[width=0.8\columnwidth]{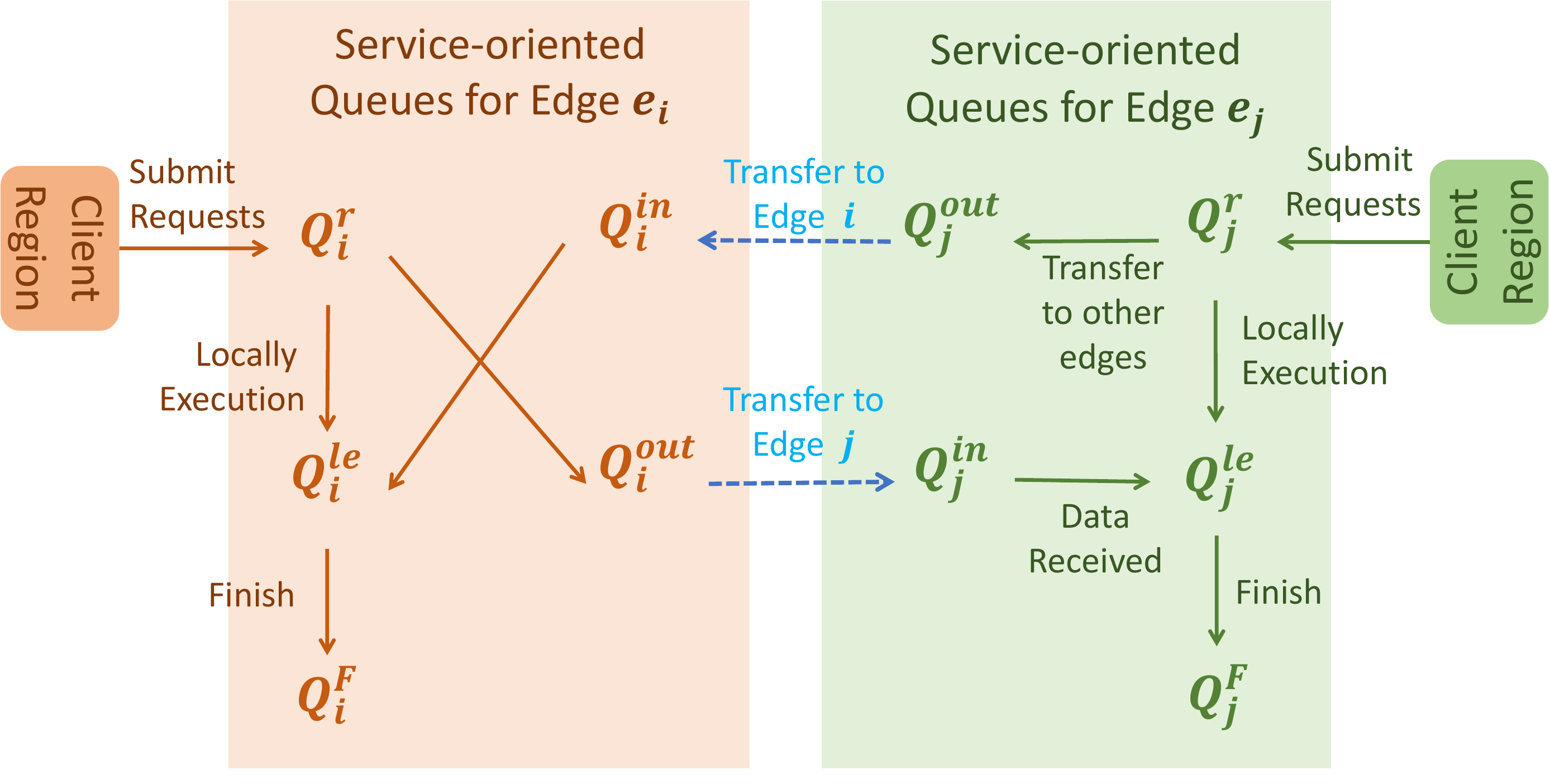}
	\caption{Illustration of the state transition of requests across queues. Clients submit requests to $Q^r$. After receiving the scheduling decision, requests that will be locally executed are transferred from $Q^r$ to $Q^{le}$, and requests that will be executed on other edges are transferred from $Q^r$ to $Q^{out}$. The requests that are transferred from other edges are saved in $Q^{in}$. Once the related data of requests in  $Q^{in}$ are received, the requests are transferred from $Q^{in}$ to $Q^{le}$. If the requests are completed, then they are transferred from $Q^{le}$ to $Q^{F}$.}
	\label{fig:QueueTrans}
\end{figure}
We design service-oriented queue models for edge $e_i$ to save requests that are at different \textcolor{black}{status}, including $Q_i^r$ for requests that are waiting for scheduling, $Q_i^{le}$ for requests that will be executed locally , $Q_i^{out}$ for requests that will be transmitted to other edges, $Q_i^{in}$ for requests that will be transferred in from other edges and $Q_i^{F}$ for requests that have been completed. The state transition of requests across queues is presented in Fig.~\ref{fig:QueueTrans}. 

When evaluating the workload of edge $e_i$, we focus on the requests in $Q_i^{le}$ and $Q_i^{in}$, \textcolor{black}{because only requests in these two queues will make use of the local resources of edge $e_i$.  
Five features are introduced to evaluate the workloads, including required computing time to complete requests in $Q_i^{le}$ (referred to as $c_i^{le}$), required data transmission time for requests in $Q_i^{in}$ (referred to as $t_i^{in}$), required computing time to complete requests in $Q_i^{in}$ (referred to as $c_i^{in}$), required results transmission time for requests in $Q_i^{le}$ (referred to as $b_i^{le}$) and required results transmission time for requests in $Q_i^{in}$ (referred to as $b_i^{in}$). }

\textcolor{black}{
The five features can be evaluated in customized mathematical approximation models. In this paper, we compute them by \eqref{eq:cle}-\eqref{eq:out-qin} respectively. 
$r$ refers to a request. $f_r$ refers to the data size of request $r$ and $u_r$ denotes data volume of $r$'s results.  $\varpi_r$, $e_q$ and $\delta_r$ represent the source edge, the execution edge and the predicted edge of request $r$, respectively. $\psi(e_q, \varpi_r)$ computes the distance between $e_q$ and $\varpi_r$, while $\psi(e_q, \delta_r)$ obtains the distance between $e_q$ and $\delta_r$. 
$C_t$ is a constant to represent the transmission speed for unit data through unit distance.} 

\begin{align}
    & c_i^{le} = \frac{\sum_{r \in Q_i^{le}} \phi_i(f_r)}{\zeta_i} \label{eq:cle} \\
    & t_i^{in} = \max \limits_{r \in Q_i^{in}} C_t f_r \psi(e_q, \varpi_r)\ \label{eq:tin} \\
    & c_i^{in} = \frac{\sum_{r \in Q_i^{in}} \phi_i(f_r)}{\zeta_i} \label{eq:cin} \\
    & b_i^{le} = \max \limits_{r \in Q_i^{in}} C_t u_r \psi(e_q, \delta_r) \label{eq:out-qle} \\
    & b_i^{in} = \max \limits_{r \in Q_i^{in}} C_t u_r \psi(e_q, \delta_r) \label{eq:out-qin}
\end{align}

In \eqref{eq:cle} and \eqref{eq:cin}, we average the computation time required to complete all tasks across multiple copies. 
\textcolor{black}{ As for \eqref{eq:tin}, \eqref{eq:out-qle} and \eqref{eq:out-qin}}, we make two assumptions based on experience to predict required data transmission time. Firstly, the data transmission time is positively correlated with both data size and transmission distance. Secondly, edges can simultaneously receive data sent by other edges from different ports. 

Evaluating the workload before each scheduling operation allows us to obtain the real-time service capacity of edge $e_q$. This real-time system state knowledge is instrumental in making well-informed scheduling decisions.

\begin{table}[t!]
	\caption{\textcolor{black}{The definitions of primary notations}}
	\renewcommand{\arraystretch}{1.2}
	\begin{center}
		\setlength{\tabcolsep}{3pt}
		\begin{tabular}{l|p{6.5cm}}
			\toprule
	Notation & Definition \\ \midrule
        $SR_k$ & The service $s_k$ oriented multi-edge cooperative subsystem. \\
        $r_z$ & The request $z$ \\
        $e_q$ & The edge $q$ \\
        $s_k$ & The service $k$ \\
        $l_{zq}$ & Whether $r_z$ is located at $e_q$. $l_{zq} \in \{0, 1\}$\\ 
        $x_{zq}$ & Whether $r_z$ is dispatched to $e_q$. $x_{zq} \in \{0, 1\}$\\
        $f_z$ & The input data size of $r_z$. \\
        $u_z$ & The output data size of $r_z$. \\
        $\phi_q(f_z)$ & The computation time to deal with data of $r_z$ at $e_q$. \\
        $p_q$ & The replica number of $s_k$ at $e_q$\\
        $c_q^{le}$ & The computation time to complete backlogs in $Q_q^{le}$.\\
        $c_q^{in}$ & The computation time to complete backlogs in $Q_q^{in}$.\\
        $\psi(e_q, e_n)$ & The transmission distance $e_n$ and $e_q$.\\
        $C_t$ & A constant to represent the transmission speed for unit data through unit distance. \\
        $t_q^{in}$ & The remaining transmission time of backlogs in $Q_q^{in}$. \\
        $[Z]$ & \{1, 2, ..., Z\}. Z denotes the number of requests. \\
        $[Q]$ & \{1, 2, ..., Q\}. Q denotes the number of edges in one control region.\\
        $[N]$ & \{1, 2, ..., N\}. N denotes the number of edges in the whole $SR_k$.\\
			\bottomrule
		\end{tabular}
		\label{table:notations}
	\end{center}
\end{table}

\subsection{Problem Formulation} \label{section:formulation}
\textcolor{black}{Each service $s_k$ oriented subsystem $SR_k$ has multiple control regions. Each control region can be formulated as $CoMEC = (\mathcal{E}, \mathcal{V}, \mathcal{P}, \mathcal{I})$.}
$\mathcal{E} = \{e_q\}_{q=1}^Q$ is a set of network edges where $s_k$ is deployed, $|\mathcal{E}|=Q$ is the number of edges. 
$\mathcal{V} = \{\phi_q\}_{q=1}^Q$ is a set of functions and $\phi_q$ represents computation time estimation function of $e_q$ (explained in the section~\ref{section:performance-estimation}).  
$\mathcal{P}$ is a $Q$-dimensional vector as well. $p_{i}$ denotes the replica numbers of $s_k$ at $e_i$. 
$\mathcal{I}$ is a $N \times 3$ matrix that specifies the current workload evaluation of edges. The entry of $\mathcal{I}$ at $q^{th}$ row, denoted as $I_{q}$, represents the workload evaluation of $e_q$. $I_{q} = (c_q^{le}, c_q^{in}, t_q^{in})$ (the definitions are explained in the section~\ref{section:workload-evaluation}). 
\textcolor{black}{Additionally, we define $N$ to represent the number of edges in the whole $SR_k$. $\psi(e_q, e_n)$ is a function to obtain the data transmission distance between $e_q$ and $e_n$. }

The requests distributed in an individual $CoMEC$ can be modeled as 
$ CoR = (\mathcal{R}, \mathcal{L}, \mathcal{H}, \mathcal{F}, \mathcal{D})$. 
$\mathcal{R} = \{ r_z\}_{z=1}^Z$ is a set of requests that require $s_k$, $|\mathcal{R}| = Z$ is the number of requests. 
$\mathcal{L}$ is a $Z \times Q$ matrix, the entry $l_{zq}$ represents whether $r_z$ is located at $e_q$ before scheduling, if yes, $l_{zq} = 1$, else, $l_{zq}=0$. 
\textcolor{black}{$\mathcal{H}$ is a $Z \times N$ matrix, the entry $h_{zn}$ represents whether the client who submits $r_z$ will move to the client region of $e_n$ after $r_z$ is completed. If yes, $h_{zn} = 1$, else, $h_{zn}=0$. 
$\mathcal{F} = \{f_z\}_{z=1}^Z$ is a set that records the size of input data for all $r_z \in \mathcal{R}$. While $\mathcal{U} = \{u_z\}_{z=1}^Z$ is a set that records the size of results for all $r_z \in \mathcal{R}$. }
$\mathcal{D}$ keeps the practical related data of requests. 

Given $CoMEC$ and $CoR$, let $\mathcal{X} \in \{0, 1\}^{Z \times Q}$ be a permutation matrix. For $\forall$ $x_{zq} \in \mathcal{X}$, $x_{zq} = 1$ represents $r_z$ is dispatched to $e_q$. 
Then the objective function of multi-edge cooperative scheduling problem can be formulated as \eqref{eq:ilp1}. The parameter $T_q$ in \eqref{eq:ilp1} denotes the required time to complete all requests that are scheduled to $e_q$ as $\mathcal{X}$. \eqref{eq:ilp1} indicates that the purpose is to get a $\mathcal{X}$ that can minimize the response time over all edges and all requests. 
\textcolor{black}{$T_q$ in \eqref{eq:ilp1} can be computed by \eqref{eq:ilp2}-\eqref{eq:ilp8} with constraints \eqref{eq:ilp9} and \eqref{eq:ilp11}. The definitions of primary notations in the formulation are summarized in Table~\ref{table:notations}. }

\begin{align}
\label{eq:ilp1}
obj. & ~~\min \limits_{\mathcal{X}}~\max \limits_{e_q \in \mathcal{E}}~~~T_q \\ 
\label{eq:ilp2}
& \mu_q = \frac{\sum_{z \in [Z]} l_{zq} x_{zq} \phi_q(f_z)}{p_q} + c_q^{le} \\
\label{eq:ilp3}
& \eta_q = \frac{\sum_{z \in [Z]} (1-l_{zq}) x_{zq} \phi_q(f_z)}{p_q} + c_q^{in} \\
\label{eq:ilp4}
& \upsilon_q^1 = \max \limits_{z \in [Z]}~ x_{zq}f_z (\sum_{m \in [Q]} l_{zm} \psi(e_m, e_q)) \\
\label{eq:ilp5}
& \kappa_q^1 = \max (C_t\upsilon_q^1, t_q^{in}) \\
\label{eq:ilp6}
& \upsilon_q^2 = \max_{z \in [Z]} x_{zq} u_z (\sum_{n \in [N]} h_{zn} \psi(e_q, e_n)) \\
\label{eq:ilp7}
& \kappa_q^2 = \max (C_t\upsilon_q^2, b_q^{in}) \\
\label{eq:ilp8}
& T_q = \max (\kappa_q^1, \mu_q) + \max (\eta_q + \kappa_q^2, b_q^{in})\\
\label{eq:ilp9}
s.t. ~
& \sum_{q \in [Q]} x_{zq} = 1 ,   \forall z \in [Z] \\
\label{eq:ilp11}
& x_{zq} \in \{0, 1\}, \forall z \in [Z], \forall q \in [Q]
\end{align}

To be specific, 
\eqref{eq:ilp2} predicts the required computing time to complete all requests that will be executed locally, including previous backlogs and new requests scheduled as $\mathcal{X}$. 
\eqref{eq:ilp3} evaluates the required computing time to complete all requests that are transferred from other edges to $e_q$, taking into account both backlogs and new requests scheduled as $\mathcal{X}$. 
\textcolor{black}{\eqref{eq:ilp4} predicts the required longest transmission time for data of new requests that are transferred from other edges to $e_q$ as $\mathcal{X}$, since multiple edges transmitting data to the same edge in parallel is allowed in our multi-edge cooperative computing system. 
Considering that the data of backlogs may haven't been transmitted to $e_q$, we design \eqref{eq:ilp5} to estimate the maximum transmission time for all requests that are transferred from other edges to $e_q$, including both backlogs and new requests scheduled as $\mathcal{X}$, through $\max$ operation over $C_t v_q$ and $t_q^{in}$. 
\eqref{eq:ilp6} and \eqref{eq:ilp7} are used to estimate the longest transmission time from the execution edge $e_q$ to the predicted edge $e_n$ for requests that are processed at $e_q$, including new requests and backlogs. 
Furthermore,  due to data transmission and local computing can run simultaneously, and the requests in $Q_q^{in}$ cannot be executed until all relevant data has been received by $e_q$, we design \eqref{eq:ilp8} to predict the required response time to complete all requests that are scheduled to $e_q$ as $\mathcal{X}$, including backlogs and new requests as well. Here $\max$ operation is used to select the case that consumes longer time.} 
While computing a feasible $\mathcal{X}$, constrains \eqref{eq:ilp9} and \eqref{eq:ilp11} must be met to ensure all requests can be scheduled to only one available edge.

Throughout the system's operation, there will be multiple task scheduling rounds. In each scheduling round, the formulation can be utilized to obtain the optimal solution $\mathcal{X}$ based on the current system state $CoMEC$ and request state $CoR$. 

\section{Lightweight Real-time Scheduler Design} \label{section:CoRaiS}
The scheduling problem has been represented as an integer linear programming formulation. There have been some solvers, such as \textit{Gurobi} and \textit{Cplex}, which can accurately solve such problems. However, the search space of multi-edge cooperative scheduling problem is $Q^Z$, and it will significantly grow as the number of edges or requests increases. Therefore, getting the optimal solution is theoretically time-consuming. It is necessary to design a novel algorithm that can provide a high-quality solution within a short and predictable time. This paper proposes a lightweight attention-based scheduler called \textit{CoRaiS}, and combines it with reinforcement learning to automatically learn a great policy that helps produce a high-quality scheduling decision in real time.

\begin{figure}[t]
	\centering
	\includegraphics[width=0.8\columnwidth]{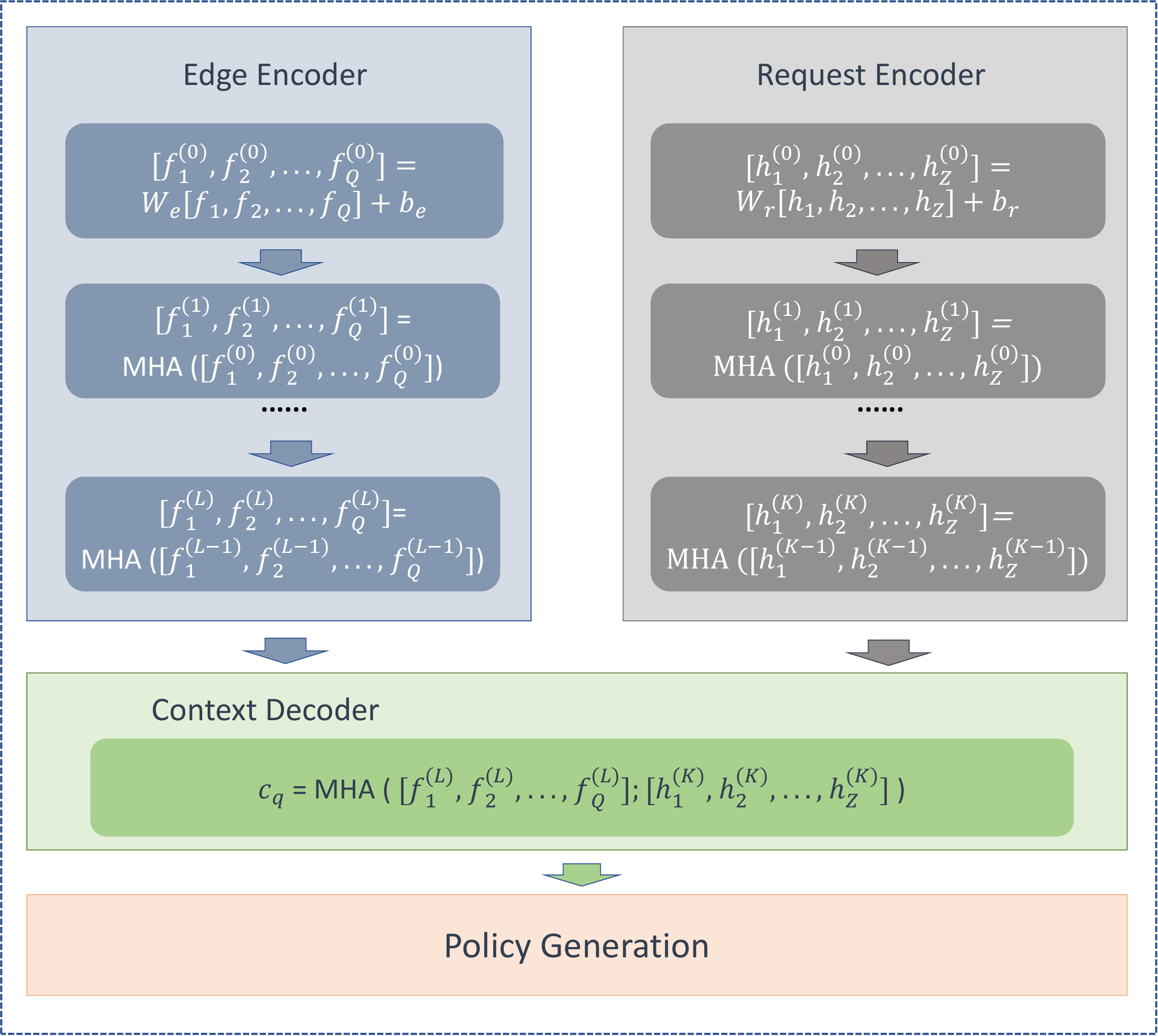}
	\caption{\textcolor{black}{Matching-on-demand architecture of \textit{CoRaiS}}}
	\label{fig:mod-architecure}
\end{figure}

\subsection{Architecture  of \textit{CoRaiS}}
\textit{CoRaiS} adopts matching-on-demand (MoD) architecture that consists of two alignment modules (edge encoder and request encoder) and one matching module (context decoder), as presented in Fig.~\ref{fig:mod-architecure}. The alignment modules are used to align specific features of heterogeneous edges through multi-dimensional information exchange. Specifically, the edge encoder embeds and aligns the performance information of edges through multi-head attention mechanism (MHA), and captures the service capacity of multi-edge system by max pooling; the request encoder has similar function with edge encoder, but works on the requests contexts, i.e. request encoder focuses on capturing and aligning the requirements of requests through MHA, and mastering the global request features by max pooling.
The matching module (context decoder) associates the capacity of multi-edge computing system with the requirements of requests through aggregating edges embeddings and requests embeddings, and produces scheduling policies to realize edge matching based on the demands of request. 

\textbf{Edge encoder:} It is a module to embed edge features. At the beginning of training, input features of edges $\mathbf{f}$ are initialized based on the current edge states. To be specific, the input features $\mathbf{f}_q$ of $e_q$ include \textit{(\romannumeral1)} coordinates $(x_q, y_q)$; \textit{(\romannumeral2)} the coefficients of $\phi_q$ and replica numbers $\zeta_q$; 
\textcolor{black}{\textit{(\romannumeral3)} workload evaluation vector $I_q$, and $I_q=(c_q^{le}, c_q^{in}, t_q^{in}, b_q^{le}, b_q^{in})$. }
The encoder computes initial $d_h$-dimensional edge embeddings $\mathbf{f}_q^{(0)}$ through a learnable linear projection with parameters $\mathbf{W}_e$ and $\mathbf{b}_e$: $\mathbf{f}_q^{(0)} = \mathbf{W}_e \mathbf{f}_q + \mathbf{b}_e$. Then the embeddings are updated through $L$ attention layers, which is motivated by Transformer \cite{vaswani2017attention} and Attention Model \cite{kool2018attention}. Each layer consists of two sublayers: a multi-head attention layer (MHA) and \textcolor{black}{an edge-wise} fully connected layer (FC). A skip-connection and batch normalization (BN) are also used at each sublayer. The operations are formulated as \eqref{eq:edge-mebdding}, where 
$\mathbf{f}_q^{(l)}$ denotes \textcolor{black}{the produced edge embedding by layer} $l \in \{1, .., L\}$.
\begin{equation}
\begin{split}
\mathbf{f}_q^{'(l)} &= \mathbf{BN}^l ( \mathbf{f}_q^{(l-1)} + \mathbf{MHA}_e^l (\{\mathbf{f}_q^{(l-1)}\}_{q=1}^{Q} )) \\
\mathbf{f}_q^{(l)} &= \mathbf{BN}^l(\mathbf{f}_q^{'(l)}, \mathbf{FC}_e^l(\mathbf{f}_q^{'(l)})
\end{split}
\label{eq:edge-mebdding}
\end{equation}	

\textbf{Request Encoder: } 
It has similar architecture and operations with \textit{edge encoder} to embed request features. But the learnable parameters are different.  The initial feature $h_z$ of $r_z$ includes 
\textcolor{black}{
\textit{(\romannumeral1)} coordinates of the source edge of $r_z$; \textit{(\romannumeral2)} input data size of $r_z$; \textit{(\romannumeral3)} coordinates of the predicted edge of $r_z$; \textit{(\romannumeral4)} output data size of $r_z$.} The initial features $\mathbf{h}_z^{(0)}$ are initialized to a $d_r$-dimensional embeddings by linear projection as \eqref{eq:requests-mebdding-initialize}. Following that, $K$ attention layers are used to update request embeddings. The operations are presented as \eqref{eq:requests-mebdding}, where $MHA_r^k$ and $FC_r^k$ are the parameters that are used to embed requests features. The operations of $\mathbf{MHA}_r^k$ are similar with them in $\mathbf{MHA}_e^l$. 
\begin{equation}
\mathbf{h}_z^{(0)} = \mathbf{W}_r \mathbf{h}_z + \mathbf{b}_r
\label{eq:requests-mebdding-initialize}
\end{equation}
\begin{equation}
\begin{split}
\mathbf{h}_z^{'(k)} &= \mathbf{BN}^k ( \mathbf{h}_z^{(k-1)} + \mathbf{MHA}_r^k (\{\mathbf{h}_z^{(k-1)}\}_{z=1}^{Z} )) \\
\mathbf{h}_z^{(k)} &= \mathbf{BN}^k(\mathbf{h}_z^{'(k)}, \mathbf{FC}_r^k(\mathbf{h}_z^{'(k)})
\end{split}
\label{eq:requests-mebdding}
\end{equation}

\textbf{Context decoder:} The system context comes from the edge embeddings $\{\mathbf{f}_q^{(L)}\}_{q=1}^{Q}$ and request embeddings $\{\mathbf{h}_z^{(K)}\}_{z=1}^{Z}$. Inspired by \cite{qi2017pointnet,xu2018powerful,hu2021bidirectional}, this paper captures global features $\hat{\mathbf{f}}$ and $\hat{\mathbf{h}}$ by max pooling operation over $\{\mathbf{f}_q^{(L)}\}_{q=1}^{Q}$ and $\{\mathbf{h}_z^{(K)}\}_{z=1}^{Z}$, respectively. Then the new context embedding $c'_q$ is produced by $\mathbf{MHA}_c$ which has $M$ heads. 
The computation process through $\mathbf{MHA}_c$ is shown as \eqref{eq:context-mebdding-attention}, where $[.,.,.]$ is the horizontal concatenation operator, $d_y = \frac{d_r}{M} = \frac{d_h}{M}$, $\mathbf{c'}_q$ is the embedding after single attention, and $\mathbf{c}_q$ is the final multi-head attention value for context embedding.
\begin{equation}
\begin{split}
& \mathbf{f}_{(c)q} = [\hat{\mathbf{f}}, \hat{\mathbf{h}}, \mathbf{f}_q^{(L)}] \\ 
& \mathbf{x}_q = \mathbf{W}_c^x \mathbf{f}_{(c)q};~
~~ \mathbf{y}_z = \mathbf{W}_c^y ~\mathbf{h}_z^{(K)}; ~
~~ \mathbf{v}_z = \mathbf{W}_c^v ~\mathbf{h}_z^{(K)} \\
& u_{qz} = \frac{ \mathbf{x}_q^T \mathbf{y}_z }{\sqrt{d_y}}; ~~ a_{qz} = \frac{e^{u_{qz}}}{\sum_{z=1}^{Z} e^{u_{qz}}}\\
& \mathbf{c'}_q = \sum_{z=1}^{Z} a_{qz} \mathbf{v}_z; ~~\mathbf{c}_q = \sum_{i=1}^{M} ~\mathbf{W}_{c,i}~\mathbf{c'}_q^i
\end{split}
\label{eq:context-mebdding-attention}
\end{equation}

The $\mathbf{MHA}$ operation in \textit{context embedding} is similar with it in \textit{edge encoder} and \textit{request encoder}, but replacing $\mathbf{f}_{(c)q}$ with $\mathbf{f}_q^{(l-1)}$ ($\mathbf{h}_z^{(k-1)}$) in \textit{edge encoder} (\textit{request encoder}).

\textbf{Policy generation: } The policy is generated by collecting importance of edges for one request $r_z$ as \eqref{eq:policy-attention}. $imp_{qz}$ denotes the importance of $e_q$ for $r_z$. $C$ is a constant. To evaluate the probability of edges getting the privilege, \textit{softmax} is introduced over all edges for each request as \eqref{eq:policy-prob}, where $a_{qz}$ specifies the probability of $e_q$ responding to $r_z$. 
\begin{equation}
\begin{split}
& \mathbf{px}_q = \mathbf{W}_{px} ~\mathbf{c}_q;~
\mathbf{py}_z = \mathbf{W}_{py} ~\mathbf{h}_z^{(H)}; \\
& u_{qz} = \frac{ \mathbf{px}_q^T \mathbf{py}_z }{\sqrt{d_{py}}}; 
imp_{qz} = C \tanh(u_{qz})
\end{split}
\label{eq:policy-attention}
\end{equation}
\begin{equation}
a_{qz} = \frac{e^{imp_{qz}}}{\sum_{q=1}^{Q} e^{imp_{qz}}}
\label{eq:policy-prob}
\end{equation}

\subsection{Training \textit{CoRaiS}}	
The scheduling probability distribution $p_\theta(\pi)=\{a_{qz}\}_{q \in [Q], z \in [Z]}$ is produced by \textit{CoRaiS}, from which we can sample a scheduling decision $\pi$. In order to train \textit{CoRaiS}, we define the expectation of the maximum response time of all requests over edges as loss function: $\mathcal{L}(\theta | g) = \mathbb{E}_{p_\theta(\pi)}[L(\pi)]$. Given $\pi$, $L(\pi) = -\hat{u}^\pi$ is computed according to \eqref{eq:edge-rewards} and \eqref{eq:global-reward}. Firstly, a local reward $u_q$ is estimated by \eqref{eq:edge-rewards}, where
$RL_q^\pi$ denotes the set of requests to be executed locally and $RT_q^\pi$ includes requests that are transferred from other edges to $e_q$, based on the decision $\pi$.  
$Y(\varpi_m, e_q)$ is a function to compute the transmission distance between the source edge $\varpi_m$ and the execution edge $e_q$ of $r_m$. $Y(\delta_m, e_q)$ obtains the transmission distance between the predicted edge $\delta_m$ and the execution edge $e_q$ of $r_m$.
The connotation of each equation in \eqref{eq:edge-rewards} is similar with \eqref{eq:ilp2}-\eqref{eq:ilp8}. Then the global reward is formulated as \eqref{eq:global-reward}.

\begin{equation}
\begin{split}
\mu_q = & \frac{\sum_{r_m \in RL_q^\pi} \phi_q(f_m)}{\zeta_q} + c_q^{le}\\
\eta_q = & \frac{\sum_{r_m \in RT_q^\pi} \phi_q(f_m)}{\zeta_q} + c_q^{in}\\
\kappa_q = & \max \{\max \limits_{r_m \in RT_q^\pi} f_m~Y(\varpi_m, e_q), ~~t_q^{in}\}\\
\beta_q = &\max \{\max \limits_{r_m \in RT_q^\pi} u_m~Y(e_q, \delta_m), b_q^{in} \} \\
u_q^\pi = &~  -(\max(\mu_q, \kappa_q) + \max(\eta_q + \beta_q, b_q^{le}))
\end{split}
\label{eq:edge-rewards}
\end{equation}

\begin{equation}
\hat{u}^\pi = \mathop{min}\limits_{e_q \in \mathcal{E}}~~ u_q^\pi
\label{eq:global-reward}
\end{equation}

We use $\mathbf{S}$-samples batch reinforcement learning (RL) and gradient descent \cite{hu2020reinforcement} to optimize $\mathcal{L}$, since the $\mathbf{S}$-samples batch gradient descent replacing the one-sample approximation in training realizes more accurate estimation of the policy, decreases training variance and speeds up convergence. Meanwhile, to encourage \textit{CoRaiS} to sufficiently explore the huge search space, we add an extra entropy loss $H(\theta)$, computed as \eqref{eq:loss-entropy}. Then the optimization function can be formulated as \eqref{eq:loss}, where $D$ is the training data set, $C_1$ and $C_2$ are the coefficients.

\begin{equation}
H_\theta(g) = - \sum_{z=1}^{Z}\sum_{q=1}^{Q} a_{qz}(\theta) \log a_{qz}(\theta)
\label{eq:loss-entropy}
\end{equation}
 
\begin{equation}
\begin{split}
A(\pi_s) =& L(\pi_s) - \frac{1}{S} \sum_{i=1}^{S}L(\pi_i)  \\
\mathcal{L}(\theta | \mathcal{D}) = & \mathbb{E}_{g \sim D}(C_1 \sum_{s=1}^{S} \log p_\theta(\pi_s | g)  A(\pi_s) \\
&~~~~~ ~~~~    - C_2   H_\theta(g))\\
\end{split}
\label{eq:loss}
\end{equation}

\subsection{Decoding Strategies of \textit{CoRaiS}}
Two decoding strategies are proposed for \textit{CoRaiS} to generate an effective scheduling decision.  
\begin{itemize}
    \item Greedy decoding: according to the generated policy of \textit{CoRaiS}, the best execution edge for each request is always selected. That is, for request $r_z$, its execute edge $e_q$ is selected by $q = \arg \max_k \{a_{kz}\}_{k=1}^Q$. 
    \item Sampling decoding:  for request $r_z$, multiple edge selections can be sampled based on multinomial probability distribution over the policy $ \{a_{qz}\}_{q=1}^Q$, and the best one is reported.  A complete scheduling decision entails execution edge selections of requests in the multi-edge computing system. 
\end{itemize}

\section{Simulation Evaluation}\label{section:simulation}
To demonstrate that \textit{CoRaiS} is able to learn a strong policy and give a real-time decision for the multi-edge scheduling problem, we designs three experiments: conventional  test, generalization test, and characteristic validation. 
The conventional test refers to evaluating the performance of \textit{CoRaiS} on a dataset that matches the same scale\footnote{In this paper, the \textit{scale} of dataset refers to the size of the multi-edge cooperative system, including the number of edges in the system and the number of requests submitted to it.} as the one it was trained on.
The generalization test refers to evaluating the performance of \textit{CoRaiS} on a dataset that has large scales than the one it was originally trained on. 
The characteristic validation is used to assess whether \textit{CoRaiS} can perceive the workload and heterogeneity of edges, and autonomously implement load balancing. 

\begin{table*}
	\caption{\textcolor{black}{Normal Test Results \\
        \textit{CoRaiS} is trained and tested on the same small-scale instances $\{(10, 5, 50), (10, 5, 100), (20, 10, 50), (20, 10, 100)\}$. \\
	\textit{Gap-M}, \textit{Time-M} and \textit{Time-S}: the \textbf{lower} the \textbf{better}.}
	}
	\renewcommand{\arraystretch}{1.0}
	\begin{center}
		\setlength{\tabcolsep}{3pt}
		\begin{tabular}{c|ccc|ccc|ccc|ccc}
		\toprule
                \multicolumn{1}{c|}{} & \multicolumn{3}{c|}{(10, 5, 50)} & \multicolumn{3}{c|}{(10, 5, 100)} & \multicolumn{3}{c|}{(20, 10, 50)} & \multicolumn{2}{c}{(20, 10, 100)} \\ 
                Methods &  Gap-M &  Time-M &  Time-S &  Gap-M &  Time-M &  Time-S &  Gap-M &  Time-M &  Time-S &  Gap-M &  Time-M &  Time-S \\ \midrule
                Gurobi(10s) & 1.0000  & 0.4294  & 1.1454  & 1.0000  & 1.3126  & 1.9815  & 1.0000  & 0.1970  & 0.0170  & 1.0000  & 0.5791  & 1.3249  \\ \midrule
                Local & 1.9602  & \textbf{0.0049}  & 0.0008  & 2.7110  & \textbf{0.0047}  & 0.0006  & 1.5214  & \textbf{0.0085}  & 0.0008  & 2.0253  & \textbf{0.0081}  & 0.0015  \\ 
                Predicted & 1.9752  & 0.0090  & 0.0015  & 2.7182  & 0.0123  & 0.0013  & 1.5268  & 0.0125  & 0.0011  & 2.0256  & 0.0155  & 0.0028  \\
                Random(1) & 2.0540  & 0.0049  & 0.0008  & 2.7707  & 0.0048  & 0.0005  & 1.5706  & 0.0086  & 0.0008  & 2.0943  & 0.0082  & 0.0015  \\ \
                Random(100) & 1.4199  & 0.0053  & 0.0008  & 1.9525  & 0.0053  & 0.0006  & 1.1211  & 0.0091  & 0.0008  & 1.4544  & 0.0089  & 0.0016  \\ 
                Random(1k) & 1.2962  & 0.0066  & 0.0011  & 1.7691  & 0.0079  & 0.0011  & 1.0624  & 0.0104  & 0.0008  & 1.3342  & 0.0112  & 0.0020  \\ \midrule
                FC1-CoRaiS(greedy) & 1.4953  & 0.0135  & 0.0019  & 1.5857  & 0.0132  & 0.0016  & 1.4443  & 0.0171  & 0.0015  & 1.5014  & 0.0163  & 0.0029  \\ 
                FC2-CoRaiS(greedy) & 1.1738  & 0.0151  & 0.0034  & 1.3871  & 0.0145  & 0.0015  & 1.0504  & 0.0175  & 0.0025  & 1.1479  & 0.0175  & 0.0031  \\ 
                FC3-CoRaiS(greedy) & 1.1802  & 0.0118  & 0.0017  & 1.3422  & 0.0115  & 0.0012  & 1.0345  & 0.0150  & 0.0018  & 1.1279  & 0.0147  & 0.0025  \\
                CoRaiS(greedy) & \textbf{1.0476}  & 0.0159  & 0.0022  & \textbf{1.0823}  & 0.0155  & 0.0015  & \textbf{1.0182}  & 0.0193  & 0.0016  & \textbf{1.0409}  & 0.0187  & 0.0031  \\ \midrule
                FC1-CoRaiS(100) & 1.2465  & 0.0138  & 0.0019  & 1.3366  & 0.0134  & 0.0016  & 1.2552  & 0.0170  & 0.0015  & 1.2291  & 0.0165  & 0.0029  \\ 
                FC2-CoRaiS(100) & 1.0544  & 0.0150  & 0.0034  & 1.2422  & 0.0146  & 0.0015  & 1.0072  & 0.0181  & 0.0025  & 1.0266  & 0.0179  & 0.0031  \\ 
                FC3-CoRaiS(100) & 1.0566  & 0.0120  & 0.0017  & 1.1949  & 0.0117  & 0.0012  & 1.0038  & 0.0152  & 0.0018  & 1.0259  & 0.0150  & 0.0025  \\ 
                CoRaiS(100) & \textbf{1.0160}  & 0.0157  & 0.0023  & \textbf{1.0353}  & 0.0154  & 0.0016  & \textbf{1.0000}  & 0.0192  & 0.0016  & \textbf{1.0051}  & 0.0185  & 0.0031  \\ \midrule
                FC1-CoRaiS(1k) & 1.1957  & 0.0138  & 0.0019  & 1.2757  & 0.0135  & 0.0016  & 1.2229  & 0.0166  & 0.0015  & 1.1805  & 0.0166  & 0.0029  \\ 
                FC2-CoRaiS(1k) & 1.0426  & 0.0151  & 0.0034  & 1.2167  & 0.0146  & 0.0015  & 1.0030  & 0.0183  & 0.0025  & 1.0193  & 0.0179  & 0.0031  \\
                FC3-CoRaiS(1k) & 1.0446  & 0.0122  & 0.0017  & 1.1750  & 0.0117  & 0.0012  & 1.0019  & 0.0156  & 0.0018  & 1.0178  & 0.0151  & 0.0025  \\ 
                CoRaiS(1k) & \textbf{1.0113}  & 0.0159  & 0.0022  & \textbf{1.0274}  & 0.0153  & 0.0018  & \textbf{1.0000}  & 0.0192  & 0.0018  & \textbf{1.0036}  & 0.0185  & 0.0032 \\ \midrule
                Gurobi(0.02s) &  (0.8\%) & 0.0772  & 0.0033  &  (0\%) &  - &  - &  (0\%) &  - &  - &  (0\%) &  - &  - \\ 
                Gurobi(0.05s) & (78.61\%)  & 0.1064  & 0.0093  &  (3.2\%) & 0.1760  & 0.0162  & (2.81\%)  & 0.1713  & 0.0092  &  (0\%) &  - &  - \\ 
                Gurobi(1s) & 1.0001  & 0.2489  & 0.3041  & 1.0015  & 0.5938  & 0.4399  & 1.0000  & 0.2115  & 0.0176  & 1.0005  & 0.4231  & 0.1432  \\ 
                Gurobi(5s) & 1.0000  & 0.3963  & 0.8883  & 1.0000  & 1.1848  & 1.4795  & 1.0000  & 0.1991  & 0.0172  & 1.0000  & 0.4875  & 0.6733 \\ \bottomrule
		\end{tabular}
		\label{table:NormalTestResult}
	\end{center}
\end{table*}

\subsection{Experiment Descriptions}
\textbf{Instance generation~~} 
The training and testing datasets are generated as the same rules.  \textcolor{black}{Given the number of edges in the whole multi-edge cooperative computing system, denoted as $N$, the number of edges in an individual control region $Q (Q < N)$ and the number of requests $Z$, } for each edge $e_q (q \in \{1, ..., N\})$, the coordinates are randomly sampled under the uniform distribution in $(0,1)^2$; the supported maximum service replica number $rn_q$ is randomly sampled in $\{1,2,3,4\} $. $C_t=3$. The functional relationship between computing time and the size of input data packets is modeled as linear functions\footnote{Without loss of generality, we model the relationship using the linear functions. But in practical, other relationships, such as quadratic function, are allowed to train \textit{CoRaiS}.} for all edges during simulations. Moreover, to represent the heterogeneity across edges, different coefficients are randomly sampled from a uniform distribution within the range $(0, 1)$. 

To make \textit{CoRaiS} can learn to adopt to any initial system-level states, we randomly generate some requests as backlogs for each edge while generating training datasets and testing datasets. The numbers of backlogs in $Q_q^{le}$ and $Q_q^{in}$ of edge $e_q$ are randomly sampled from $(0, 100)$.
\textcolor{black}{For any backlog $r_x$ in $Q_q^{le}$, its input data size $f_x$ is uniformly sampled from $(0,1)$, output data size $u_x$ is uniformly sampled from $(0, 0.1)$, and its predicted edge $\delta_x$ is sampled from $[N]$. For any backlog $r_k$ in $Q_q^{in}$, its source edge $\varpi_k$ is sampled from $[Q]-\{q\}$, its predicted edge $\delta_k$ is sampled from $[N]$, and its input and output data sizes are uniformly sampled from $(0,1)$ and $(0, 0.1)$}. With these backlogs, the service-oriented workload evaluation is carried out as \eqref{eq:cle}-\eqref{eq:cin}.

For any new request $r_z$ that will be scheduled in current period, its source edge $\varpi_z$ is sampled from $[Q]$, \textcolor{black}{its predicted edge $\delta_z$ is sampled from $[N]$, and the related input data size $f_z$ and output data size $u_z$ are uniformly sampled from $(0,1)$ and $(0, 0.1)$, respectively.}

\textbf{Hyperparameters~~} 
Learnable parameters are initialized as Uniform($-1/\sqrt{d}, 1/\sqrt{d}$), with $d$ is the input dimension; learning rate $lr=1e-5$; batch size is 128, $S=64$ while using $S$-samples batch RL; $C_1=10$ and $C_2=0.5$. 
The \textit{edge encoder} and \textit{request encoder} have $L=5$ and $K=3$ attention layers, respectively. $\mathbf{MHA}$ and $\mathbf{FC}$ in two embedding modules have same structure. $\mathbf{MHA}$ has 8 heads, $\mathbf{FC}$ has one hidden sublayer with dimension 512 and ReLu activation. 

\textit{CoRaiS} is trained from 40000 batches on the device with $2 \times$Intel(R) Xeon(R) Gold 5215 CPU that can provide 40 processors and $2 \times$NVIDIA GTX 2080Ti, and \textit{CoRaiS} uses Adam optimizer, PyTorch 1.11 framework and python 3.8 on Ubuntu 18.04.

\textbf{Baselines~~} 
It must be mentioned that as our best knowledge, \textit{CoRaiS} is the first artificial intelligence based work that breaks many assumptions of prior works, deals with a multi-edge computing system where every edge is allowed to receive requests from clients, and optimizes an objective function that minimize the responding time over all requests. In order to evaluate the performance of \textit{CoRaiS}, the results generated by \textit{CoRaiS} are compared with that produced by the exact solver \textit{Gurobi}. \textit{Gurobi} is able to calculate exact solutions. Meanwhile, we propose three heuristic approaches (\textit{Local}, \textit{Random} and \textit{Predicted}) as the lightweight scheduling baselines in the scenarios. Three learning-based baselines are designed as well. 
 
\begin{itemize}
	\item \textit{Exact baseline}: \textit{Gurobi} is one of the state-of-the-art solver for integer linear programming problems. However, \textit{Gurobi} may fall into long-term calculation due to the huge search space of multi-edge scheduling. Therefore, it is necessary to give a computation time limitation $(x~s)$. 
	
	\item \textcolor{black}{\textit{Lightweight heuristic baselines}}: two algorithms are used to provide heuristic baselines.  \textit{(\romannumeral1)} \textit{Local}: executing all requests at their source locations.  \textit{(\romannumeral2)} \textit{Random}: randomly sampling the execution edges for all requests. It is allowed to sample multiple times and report the best one.  \textcolor{black}{\textit{(\romannumeral3)} \textit{Predicted}: executing all requests at their predicted edges, but if the predicted edges of some requests are out of the control region, the requests will be executed at their source edges. }
	
	\item \textit{Learning-based baselines (ablation studies for aligning modules of \textit{CoRais})}: three synthetic neural network models are used to illustrate the effectiveness of \textit{CoRaiS}. \textit{(\romannumeral1)} \textit{FC1-CoRaiS}: replacing the multi-head attention aligning mechanism of \textit{edge encoder} in \textit{CoRaiS} with the multi-layer perceptron, and maintaining the same number of neuron parameters.
	\textit{(\romannumeral2)} \textit{FC2-CoRaiS}: adopting the similar MoD architecture with \textit{CoRaiS}, but using the multi-layer perceptron to replace the multi-head attention aligning mechanism in \textit{request encoder}.
	\textit{(\romannumeral3)} \textit{FC3-CoRaiS}: adopting MoD structure, but without aligning mechanisms in both \textit{edge encoder} and \textit{request encoder}, only using multi-layer perceptron to embed edges' features and requests' features. 
	Since \textit{FC1-CoRaiS}, \textit{FC2-CoRaiS} and \textit{FC3-CoRaiS} adopt the similar MoD architecture and have the same input/output with \textit{CoRaiS}, they are able to use the same decoding strategies with \textit{CoRaiS} as well. 
\end{itemize}

\begin{table*}
	\caption{\textcolor{black}{
        Generalization Test Results on Medium-scale Instances. \\
        \textit{CoRaiS} is trained on small-scale instances (10, 5, 100) and directly applied into medium-scale instances. $(x\%)$ refers to $x\%$ of the instances are solved. \\
	\textit{Gap-M}, \textit{Time-M} and \textit{Time-S}: the \textbf{lower} the \textbf{better}.}
	}
	\renewcommand{\arraystretch}{1}
	\begin{center}
		\setlength{\tabcolsep}{3pt}
		\begin{tabular}{c|ccc|ccc|ccc|ccc}
			\toprule
			 \multicolumn{1}{c|}{} & \multicolumn{3}{c|}{(20, 10, 200)} & \multicolumn{3}{c|}{(60, 30, 400)} & \multicolumn{3}{c|}{(100, 50, 600)} & \multicolumn{3}{c}{(100, 50, 800)} \\ 
Methods &  Gap-M &  Time-M &  Time-S &  Gap-M &  Time-M &  Time-S &  Gap-M &  Time-M &  Time-S &  Gap-M &  Time-M &  Time-S \\ \midrule
Gurobi(10s) & 1.0000 & 2.0173  & 3.0761  & 1.0000 & 6.1590  & 1.1698  & (97.60\%) & 16.7801  & 1.2889  & (58.00\%) & 20.0647  & 0.7842  \\ \midrule
Local & 3.0285  & \textbf{0.0088}  & 0.0008  & 2.4070  & \textbf{0.0243}  & 0.0019  & 2.2841  & \textbf{0.0394}  & 0.0028  & 2.4602  & \textbf{0.0394}  & 0.0024  \\ 
Predicted & 3.0860  & 0.0265  & 0.0018  & 2.4735  & 0.0620  & 0.0032  & 2.3190  & 0.0968  & 0.0047  & 2.4820  & 0.1145  & 0.0042  \\
Random(1k) & 1.9392  & 0.0147  & 0.0013  & 1.6445  & 0.0362  & 0.0024  & 1.5753  & 0.0574  & 0.0034  & 1.7049  & 0.0608  & 0.0034  \\ \midrule
CoRaiS(greedy) & 1.0739  & 0.0206  & 0.0014  & 1.0139  & 0.0373  & 0.0020  & 0.9822  & 0.0533  & 0.0030  & 0.9400  & 0.0523  & 0.0025  \\ 
CoRaiS(100) & 1.0246  & 0.0201  & 0.0016  & 1.0002  & 0.0362  & 0.0021  & 0.9672  & 0.0523  & 0.0033  & 0.8999  & 0.0520  & 0.0028  \\ 
CoRaiS(1k) & \textbf{1.0202}  & 0.0202  & 0.0015  & \textbf{1.0002}  & 0.0357  & 0.0022  & \textbf{0.9672}  & 0.0519  & 0.0032  & \textbf{0.8999}  & 0.0514  & 0.0028 \\ \midrule
Gurobi(0.05s) &  (0\%) &  - &  - &  (0\%) &  - &  - &  (0\%) &  - &  - &  (0\%) &  - &  - \\ 
Gurobi(1s) & 1.0062  & 0.9511  & 0.2605  &  (0\%) &  - &  - &  (0\%) &  - &  - &  (0\%) &  - &  - \\ 
Gurobi(5s) & 1.0006  & 1.4622  & 1.5466  & 1.0032  & 6.0602  & 0.8031  &  (1.2\%) & 12.4700  & 0.4302  &  (0\%) &  - &  - \\ \bottomrule
		\end{tabular}
		\label{table:generalization-result}
	\end{center}
\end{table*}

\textbf{Performance indexes~~} Three indexes are used to evaluate performance. 
\begin{itemize}
    \item \textit{Gap-M}: average quality difference of solutions that are generated by \textit{CoRaiS} and other baselines, compared with the solution generated by \textit{Gurobi(10s)} in the same instance. Solutions from \textit{Gurobi(10s)} are considered as the optimal benchmark in the simulations. The \textit{gap} is computed by \eqref{eq:gap-computing}. \textcolor{black}{$\aleph$ includes \textit{CoRaiS} and other baselines. $\pi$ signifies the best solution obtained from the specific approach $b$ ($b\in \aleph$). $\hat{\pi}$ refers to the best solution generated by $Gurobi(10s)$.} \textcolor{black}{$L(\pi)$ refers to the predicted response time for all requests in the multi-edge cooperative computing system, and $L(\pi)$ is computed by \eqref{eq:global-reward}.} \textit{Gap-M} is critical because the multi-edge cooperative computing system also require the real-time approach to produce a high-quality solution. 
    
\begin{equation}
gap_b = \frac{L(\pi|b)}{L(\hat{\pi}|Gurobi(10s))}, ~~\forall b \in \aleph
\label{eq:gap-computing}
\end{equation}

    \item \textcolor{black}{\textit{Time-M}: average time taken to make scheduling decisions. \textit{Time-M} is an important index to estimate whether the methods can provide effective decision in time. }
    
    \item  \textcolor{black}{ \textit{Time-S}: the standard variance of computing times to make scheduling decisions.  \textit{Time-S} can evaluate the stability of decision-making time. }
\end{itemize}

\begin{table*}
	\caption{\textcolor{black}{
        Generalization Test Results on Lager-scale Instances\\
        \textit{CoRaiS} is trained on small-scale instances (10, 5, 100). 
	\textit{Cost-M}, \textit{Gap-M} and \textit{Time-M}: the \textbf{lower} the \textbf{better}.}
	}
	\renewcommand{\arraystretch}{1.0}
	\begin{center}
		\setlength{\tabcolsep}{3pt}
		\begin{tabular}{c|ccc|ccc|ccc|ccc|ccc}
			\toprule
\multicolumn{1}{c|}{} & \multicolumn{3}{c|}{(200, 100, 2k)} & \multicolumn{3}{c|}{(600, 300, 4k)} & \multicolumn{3}{c|}{(1000, 500, 6k)} & \multicolumn{3}{c|}{(2k, 1k, 10k)} & \multicolumn{3}{c}{(2k, 1k, 15k)} \\ 
Methods &  Cost-M & Gap-M &  Time-M &  Cost-M & Gap-M &  Time-M &  Cost-M & Gap-M &  Time-M &  Cost-M & Gap-M &  Time-M &  Cost-M & Gap-M &  Time-M\\ \midrule
Gurobi(5s) & (0\%) & - & - & (0\%) & - & - & (0\%) & - & - & (0\%) & - & - & (0\%) & - & - \\
Gurobi(10s) & (3.1\%) & - & - & (0\%) & - & - & (0\%) & - & - & (0\%) & - & - & (0\%) & - & - \\ \midrule
CoRaiS(1k) & \textbf{5.1795}  & \textbf{1.0000}  & 0.0600  & \textbf{5.8071}  & \textbf{1.0000}  & 0.3783  & \textbf{6.0991}  & \textbf{1.0000}  & 0.9563  & \textbf{7.1192}  & \textbf{1.0000}  & 3.1611  & \textbf{8.0548}  & \textbf{1.0000}  & 4.6019  \\ \midrule
Local & 16.6937  & 3.2390  & \textbf{0.0403}  & 14.5508  & 2.5340  & \textbf{0.1285}  & 14.1081  & 2.3484  & \textbf{0.2129}  & 13.7089  & 2.0241  & \textbf{0.6159}  & 16.7721  & 2.2784  & \textbf{0.4401}  \\
Predicted & 17.0419  & 3.3059  & 0.1293  & 14.5757  & 2.5398  & 0.3452  & 14.1772  & 2.3591  & 0.4828  & 13.7813  & 2.0354  & 1.2682  & 16.7845  & 2.2807  & 1.1119  \\
Random(100) & 13.0677  & 2.5316  & 0.0441  & 12.0275  & 2.0924  & 0.1426  & 11.8956  & 1.9807  & 0.2227  & 11.6695  & 1.7260  & 0.6598  & 14.4036  & 1.9613  & 0.6160  \\
Random(500) & 12.4988  & 2.4216  & 0.0573  & 11.6141  & 2.0199  & 0.2261  & 11.5509  & 1.9227  & 0.5283  & 11.3679  & 1.6806  & 1.7379  & 14.0588  & 1.9127  & 2.4922  \\ 
Random(1k) & 12.3202  & 2.3865  & 0.0796  & 11.4478  & 1.9912  & 0.4145  & 11.4235  & 1.9008  & 1.0049  & 11.2717  & 1.6671  & 3.2203  & 13.9004  & 1.8908  & 4.5946 \\ \midrule
CoRaiS(10) & 5.2763  & 1.0186  & 0.0485  & 6.0588  & 1.0396  & 0.1547  & 6.4901  & 1.0604  & 0.2459  & 7.9634  & 1.1092  & 0.7364  & 9.2483  & 1.1436  & 0.6991  \\ 
CoRaiS(100) & 5.2053  & 1.0049  & 0.0597  & 5.9007  & 1.0139  & 0.1562  & 6.1954  & 1.0147  & 0.2724  & 7.4796  & 1.0441  & 0.7771  & 8.5213  & 1.0547  & 0.8522  \\ 
CoRaiS(500) & 5.1847  & 1.0009  & 0.0510  & 5.8259  & 1.0027  & 0.2175  & 6.1188  & 1.0035  & 0.5298  & 7.1911  & 1.0093  & 1.7753  & 8.1929  & 1.0183  & 2.6391  \\ \bottomrule

		\end{tabular}
		\label{table:generalization-result-larger}
	\end{center}
\end{table*}

\begin{table}
	\caption{\textcolor{black}{Characteristics Validation Results \\
 \textit{CoRaiS} is trained on scheduling problems (10, 5, 100). ~~~\textit{CoRaiS(1k)} is used to sample near-optimal solutions. 
		 \textit{LB}: Load Balancing; 
		 \textit{WP}: Workload Perception; 	
		 \textit{HA}: Heterogeneity Awareness;
		 \textit{EReqN}: the number of requests that are executed at each edges; 
		 \textit{LCost}: the corresponding response time.
	}
 }
	\renewcommand{\arraystretch}{1.0}
	\begin{center}
		\setlength{\tabcolsep}{3pt}
		\begin{tabular}{c|cc|cc|cc}
			\toprule
			\multicolumn{1}{c|}{} & \multicolumn{2}{c|}{LB} & \multicolumn{2}{c|}{WP} & \multicolumn{2}{c}{HA} \\ 
			No. & EReqN & LCost & EReqN & LCost & EReqN & LCost \\ \midrule
			A & 21.5950 & 4.1647 & 19.7030 & 14.7180 & 11.9850 & 4.2231 \\
			B & 19.6170 & 4.1958 & 20.2250 & 14.9676 & 14.8660 & 4.2165 \\ 
			C & 19.5190 & 4.1825 & 20.0240 & 14.8704 & 21.5160 & 4.0770 \\
			D & 19.6090 & 4.1935 & 20.0240 & 14.8709 & 25.8380 & 3.6357 \\ 
			E & 19.6600 & 4.1953 & 20.0240 & 14.8709 & 25.7950 & 2.4347 \\
			\bottomrule
			
		\end{tabular}
		\label{table:characteristic-verification}
	\end{center}
\end{table}

\subsection{Results Analysis}
\subsubsection{Analysis of Conventional Test Results} 
\textit{CoRaiS} is a lightweight model that has about 4 million learnable parameters. We trained \textit{CoRaiS} on four scales, i.e., $(N, Q, Z) \in \{(10, 5, 50), (10, 5, 100), (20, 10, 50), (20, 10, 100)\}$ where $(N, Q, Z)$ denotes the number of edges in the whole multi-edge cooperative computing system, the number of edges in an individual control region $Q$ and the number of requests $Z$.
We test the performance of learned policies on the same scale problems, and the results are shown in Table~\ref{table:NormalTestResult}.

(\romannumeral1) \textit{Gurobi} can quickly obtain the optimal solutions for small-scale problems, and the average time costs for $(N, Q, Z) \in \{(10, 5, 50), (20, 10, 50), (20, 10, 100)\}$ are less than $1s$. However, as the problem scale becomes larger, such as when $(N, Q, Z) = (10, 5, 100)$, the solving time increases significantly, even sometimes, \textit{Gurobi} spends $10s$ but only get a sub-optimal solution. 

(\romannumeral2) \textcolor{black}{Lightweight heuristic approaches (\textit{Local}, \textit{Predicted} and  \textit{Random}) take near-zero time to give a solution, but the quality is usually far away from the optimum. }

(\romannumeral3) \textcolor{black}{The solving time taken by \textit{CoRaiS(greedy, 100, 1k)} is close to $0.02s$ with low variance $0.003s$, which is less than \textit{Gurobi} in both \textit{Time-M} and \textit{Time-S}. } The gap closing to 1 shows \textit{CoRaiS} successfully learns an efficient policy that produces high-quality solutions. Therefore, \textit{CoRaiS has potential to provide real-time and high-quality decision to support efficient operations of the multi-edge cooperative computing system.}

(\romannumeral4) \textit{CoRaiS} and other three learning-based models (\textit{FC1-CoRaiS}, \textit{FC2-CoRaiS} and \textit{FC3-CoRaiS}) adopt the same decoding strategies (greedy, sampling).  The comparison results show that   \textit{the two alignment mechanisms of edge/request features play important roles in promoting policy learning. }

(\romannumeral5) \textcolor{black}{Because \textit{CoRaiS} spends near $0.02s$ on solving the problems, we explore the performance of \textit{Gurobi(0.02s)} as well. 
The experimental results illustrate that it is very difficult for \textit{Gurobi} to solve the problem within such a short time, even to calculate an approximate solution. Then we relax the time constraints to $0.05s$, $1s$ and $5s$. The results generated by 
\textit{Gurobi(0.05 s)} show that double relaxation does not improve performance much. 
The proportion of problems that are not solved approximately is still high. Only some simple problems can obtain near-optimal solutions. The results generated by \textit{Gurobi(1s)} and \textit{Gurobi(5s)} show 50X and 200X relaxation help a lot, since all problems get their sub-optimal solutions. However, the average decision-making time is too long  for a real-time computing system, and the stability of time usage also performs not well.} 

\subsubsection{Analysis of Generalization Test Results}
We train \textit{CoRaiS} on instances under problem scale setting $(10, 5, 100)$. \textcolor{black}{The learned model is directly applied into medium-scale and large-scale instances. The results are presented in Table~\ref{table:generalization-result} and Table~\ref{table:generalization-result-larger}, respectively.}

\textcolor{black}{In the medium-scale instances, the number of edges are limited to be more than 10 but less than 100, while the number of requests are constrained to several hundred levels. We present the results on four medium scales in Table~\ref{table:generalization-result}, including $(20, 10, 200), (60, 30, 400), (100, 50, 600)$ and $(100, 50, 800)$. \textit{Gurobi(10s)} takes a long time to compute a good solution, and its time cost increases as the problem scale becomes larger, i.e. from $2s$ for $(20, 10, 200)$ to $20s$ for $(100, 50, 800)$. However, as the scales become too large, \textit{Gurobi(10s)} cannot solve the problem within the limited computing time, such as instances on $(100, 50, 600)$ and $(100, 50, 800)$. }
\textcolor{black}{Compared with \textit{Gurobi(10s)}, \textit{CoRaiS} costs a shorter time $(<0.05s)$ to obtain a high-quality near-optimal solution by sampling, and its time cost does not increase significantly, from $0.02s$ for $(20, 10, 200)$ to $0.05s$ for $(100, 50, 800)$, even though the problem scale becomes 20X larger. 
We force \textit{Gurobi} to provide solutions within $0.05s$ to check whether it can get similar performance with \textit{CoRaiS}. The results presented in Table~\ref{table:generalization-result} show that \textit{Gurobi} can hardly solve large-scale problems within a short time. Then we relax the time limitation to 20X and 100X, i.e. $1s$ and $5s$, the comparison results show that only a few large-scale problems can be solved. 
Lightweight heuristic approaches are also used as baselines. The comparison results show that even though CoRaiS has a decision-making time twice that of heuristic approaches, it can produce much better assignment decisions to realize shorter response time over all requests. 
}

\textcolor{black}{
Furthermore, we conduct generalization testing on large-scale instances. The number of edges in large-scale instances is set to up to 1k, while the number of requests ranges from several thousand to 15k.  Table~\ref{table:generalization-result-larger} displays the results on five scales, including (200, 100, 2k), (600, 300, 4k), (1k, 500, 6k), (2k, 1k, 10k) and (2k, 1k, 15k). To facilitate the evaluation of different algorithms' performance, we introduce a new metric, \textit{Cost-M}, estimating the average response time over requests, measured in seconds. In the table, we observe that \textit{Gurobi(5s)} and \textit{Gurobi(10s)} struggle to handle the large-scale instances. Therefore, we select CoRaiS(1k) as the baseline and compare its results with lightweight heuristics. It's evident that decision-making time increases with problem scales for all scheduling approaches. Lightweight heuristic approaches, especially \textit{Local}, enable faster decision-making. However, their decision quality is significantly inferior to that of CoRaiS(1k). The \textit{Cost-M} of lightweight heuristics is usually more than twice that of CoRaiS(1k). Additionally, to address the need for swift and effective decision-making in specific cases, we relax the sampling times of CoRaiS to 10, 100, and 500. The results show that with fewer samples, CoRaiS can speed up decision-making while ensuring more effective scheduling than baseline methods.}


\subsubsection{Analysis of Characteristic Validation Results}

(\romannumeral1) \textbf{Load balancing (LB):} We design five homogeneous edges and push same backlogs on them, so that the initial system-level state of edges are same, and the time of edges responding to backlogs satisfies $b_E = b_D = b_C = b_B = b_A$. 
Then we conduct 10k experiments on the system. In each experiment, 100 same requests are submitted to $e_A$, and \textit{CoRaiS} is used to provide a scheduling solution. The results are presented in Table~\ref{table:characteristic-verification}(LB). 
The number of requests executed at five edges are approximately equal, and the response time are very close. \textit{Therefore, we claim that \textit{CoRaiS} learned to optimize scheduling through load balancing without manual intervention}. 

(\romannumeral2) \textbf{Workload perception (WP):} We design five homogeneous edges and induce differences in response time to backlogs by pushing different numbers of requests to each edge, so that the initial system-level state of edges are different. The time of edges responding to backlogs satisfies $b_E \leq b_D \leq b_C \leq b_B < b_A$. 
Then 10k experiments submitting 100 requests to $e_A$ are conducted. \textit{CoRaiS} makes decisions to schedule requests. As shown in Table~\ref{table:characteristic-verification}(WP), the number of requests to be dispatched to each edge is different and follows the order $n_E \geq n_D \geq n_C \geq n_B > n_A$ on average. Meanwhile, there is no significant difference in response time after scheduling. 
\textit{Therefore, we claim \textit{CoRaiS} is able to perceive workload of edges while scheduling.}  

(\romannumeral3) \textbf{Heterogeneity awareness (HA):} We design five heterogeneous edges, that is, the \textit{computation time estimation functions} of edges are different. We arrange the computing performance of edges in the order of $E>D>C>B>A$. Then we equalize the response times of the edges to backlogs by adjusting the number of requests.
After that, 10k experiments are conducted. In each experiment, 100 same requests are submitted to $e_A$, and \textit{CoRaiS} is used to provide a scheduling solution. According to the results in Table~\ref{table:characteristic-verification}(HA), the more powerful edge serves more requests, and the corresponding response time of all edges are very close. 
\textit{Therefore, we claim \textit{CoRaiS} is able to recognize heterogeneity while cooperating multi-edges.}  

\section{Prototype Evaluation}\label{section:prototype}

\begin{figure}[t]
	\centering
	\includegraphics[width=\columnwidth]{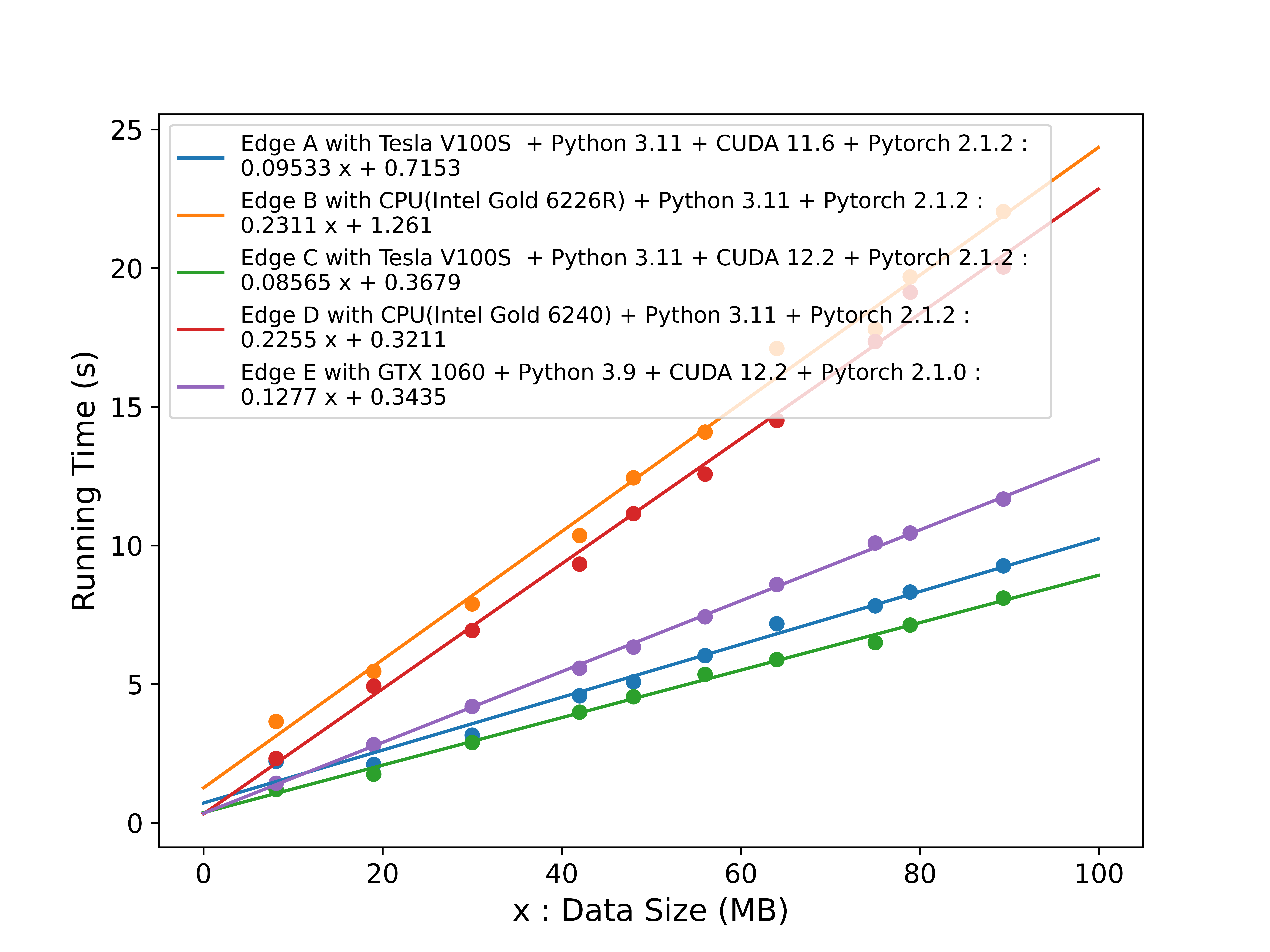}
	\caption{\textcolor{black}{The linear relationship between running time and the size of data packets for \textit{Model Soups} \cite{pmlr-v162-wortsman22a} on the five heterogeneous devices (Edge A, Edge B, Edge C, Edge D, Edge E). Points are actual statistical data, while lines are function relationships fitted based on the data. }} 
	\label{fig:prototype-computation-configuration}
\end{figure}

\subsection{Environment Description}
\textcolor{black}{Prototype experiments are conducted to illustrate the performance of proposed CoRaiS. Through out the experiments, five heterogeneous servers are deployed as edge computing nodes, and the servers are networked using a dedicated VPN. The configurations and the computation time estimation functions (\textit{Model Soups}\cite{pmlr-v162-wortsman22a} oriented) of the servers are shown in Fig.~\ref{fig:prototype-computation-configuration}. 
The number of service replicas deployed on the edges are (2, 4, 3, 3, 1). 
Three levels of requests are submitted to the edges, comprising small-size requests with 2.6MB of data, medium-size requests with 7.9MB, and large-size requests with 18.6MB. 
Five-scale experiments are established, i.e., the number of requests involves 50, 100, 200, 500, and 1000. In each experiment, the distribution percentages for these request sizes are 50\%, 30\%, and 20\%, respectively. 
The primary index collected is the physical response time, considered the most crucial metric.  }

\textcolor{black}{
Two kinds of experiments are carried out, including \textit{equal-distribution testing} and \textit{shifted-distribution testing}. 
\begin{itemize}
    \item \textit{Equal-distribution testing}: Each request's source edge and predicted edge are uniformly sampled from the five available edges, while its execution edge is determined through scheduling algorithms. The distribution setting for requests aligns with the data generation method used during the training of CoRaiS.
    \item \textit{Shifted-distribution testing}: The probabilities of each request being submitted to the five edges are predefined, i.e. [50\%, 20\%, 20\%, 5\%, 5\%], which makes the initial distribution of requests on edges be quite different from the uniform distribution. The probabilities of each request's predicted edge is predefined as well, that is, [10\%, 10\%, 40\%, 25\%, 15\%]. The execution edge of each request is decided by scheduling algorithms. 
\end{itemize}}
\textcolor{black}{
To enable real-time scheduling of requests, lightweight scheduling algorithms were utilized in the experiments. These include CoRaiS\footnote{\textit{CoRaiS} is trained on scheduling problems (10, 5, 100).} with a sampling decoding strategy, denoted as CoRaiS(1k), as well as three heuristic approaches capable of making decisions within 0.1 seconds, i.e. \textit{Local}, \textit{Predicted}, and \textit{Random(1k)}.}

\subsection{Results Analysis}
\textcolor{black}{
For \textit{equal-distribution testing}, the practical response times of requests are summarized in Table~\ref{table:prototype-uniform}. From the table, it is evident that CoRaiS(1k) consistently facilitates more effective scheduling decisions, resulting in faster response times for requests. }

\textcolor{black}{
For \textit{shifted-distribution testing}, the experimental results are summarized in Table~\ref{table:prototype-greedy}. The findings outlined in Table~\ref{table:prototype-greedy} reveal that CoRaiS is adaptable to request distributions different from those it was trained on, enabling more effective decisions to accelerate response times for requests.}

\begin{table}
	\caption{\textcolor{black}{Practical response time of requests while establishing equal-distribution testing (measured in seconds) }
 }
	\renewcommand{\arraystretch}{1.0}
	\begin{center}
		\setlength{\tabcolsep}{3pt}
		\begin{tabular}{c|cccc}
			\toprule
	ReqNum & CoRaiS(1k) & Local & Predicted & Random(1k) \\ \midrule
        50 & \textbf{6.361}  & 11.206  & 8.223  & 7.985  \\ 
        100 & \textbf{13.503}  & 14.375  & 15.891  & 15.101  \\
        200 & \textbf{26.613}  & 33.124  & 31.087  & 35.854  \\ 
        500 & \textbf{67.972}  & 72.965  & 77.099  & 80.369  \\ 
        1000 & \textbf{130.385}  & 141.639  & 146.179  & 157.577 \\ \bottomrule	
		\end{tabular}
		\label{table:prototype-uniform}
	\end{center}
\end{table}

\begin{table}
	\caption{\textcolor{black}{Practical response time of requests while establishing shifted-distribution testing (measured in seconds) }
 }
	\renewcommand{\arraystretch}{1.0}
	\begin{center}
		\setlength{\tabcolsep}{3pt}
		\begin{tabular}{c|cccc}
			\toprule
	ReqNum & CoRaiS(1k) & Local & Predicted & Random(1k) \\ \midrule
        50 & \textbf{7.454}  & 12.834  & 8.624  & 8.187  \\
        100 & \textbf{13.743}  & 22.908  & 17.056  & 16.888  \\
        200 & \textbf{27.604}  & 46.715  & 30.434  & 31.749  \\ 
        500 & \textbf{67.630}  & 115.215  & 94.336  & 78.333  \\ 
        1000 & \textbf{131.461}  & 224.968  & 176.093  & 157.254 \\ \bottomrule		
		\end{tabular}
		\label{table:prototype-greedy}
	\end{center}
\end{table}

\section{Conclusion} \label{section:conclusion}
\textcolor{black}{To facilitate the effective cooperation among edges, this paper firstly introduces the system-level state evaluation model to shield edges heterogeneity in hardware configuration and redefine the service capacity at edges. Following that, the territorial schedulers are introduced into the service-oriented multi-edge cooperative computing system, and the multi-edge scheduling problem is formulated to an integer linear programming formulation to inspire the design of multi-edge cooperative algorithms. After that, \textit{CoRaiS}, a learning-based lightweight real-time scheduler, is proposed to minimize the response time over all distributed arriving requests. The experimental results demonstrate that \textit{CoRaiS} successfully learned a strong policy to make high-quality scheduling in real time, irrespective of request arrival patterns and system scales. }

\textcolor{black}{The proposed cooperation models and algorithms still have limitations in supporting multi-edge cooperation at a global level due to the lack of support for multi-scheduler cooperation. In the future, we will continue studying multi-edge cooperative computing and plan to design a hierarchical interoperability model based on current research progresses. The model will categorize schedulers into medium-level schedulers and a top-level scheduler. Medium-level schedulers can cooperate with each other under the control of their higher-level scheduler. With the improved interoperability model, scheduling problems will be reformulated from the different perspectives of medium-level schedulers and the top scheduler. Scheduling algorithms will also be designed with consideration of the accumulative decision-making time delay across the hierarchy of schedulers. 
The updated cooperation models and algorithms aim to ensure the effective operation of a scalable multi-edge cooperative computing system, regardless of the scheduling levels and the number of edges and requests in the system. 
}

\section*{Acknowledgement}
We are grateful to New Lossless Data Center at Purple Mountain Laboratories for generously providing sufficient computing power for model training and inferences. 

\bibliographystyle{IEEEtranbst}
\bibliography{huyujiao}

\begin{IEEEbiography}[{\includegraphics[width=1in,height=1.25in,clip,keepaspectratio]{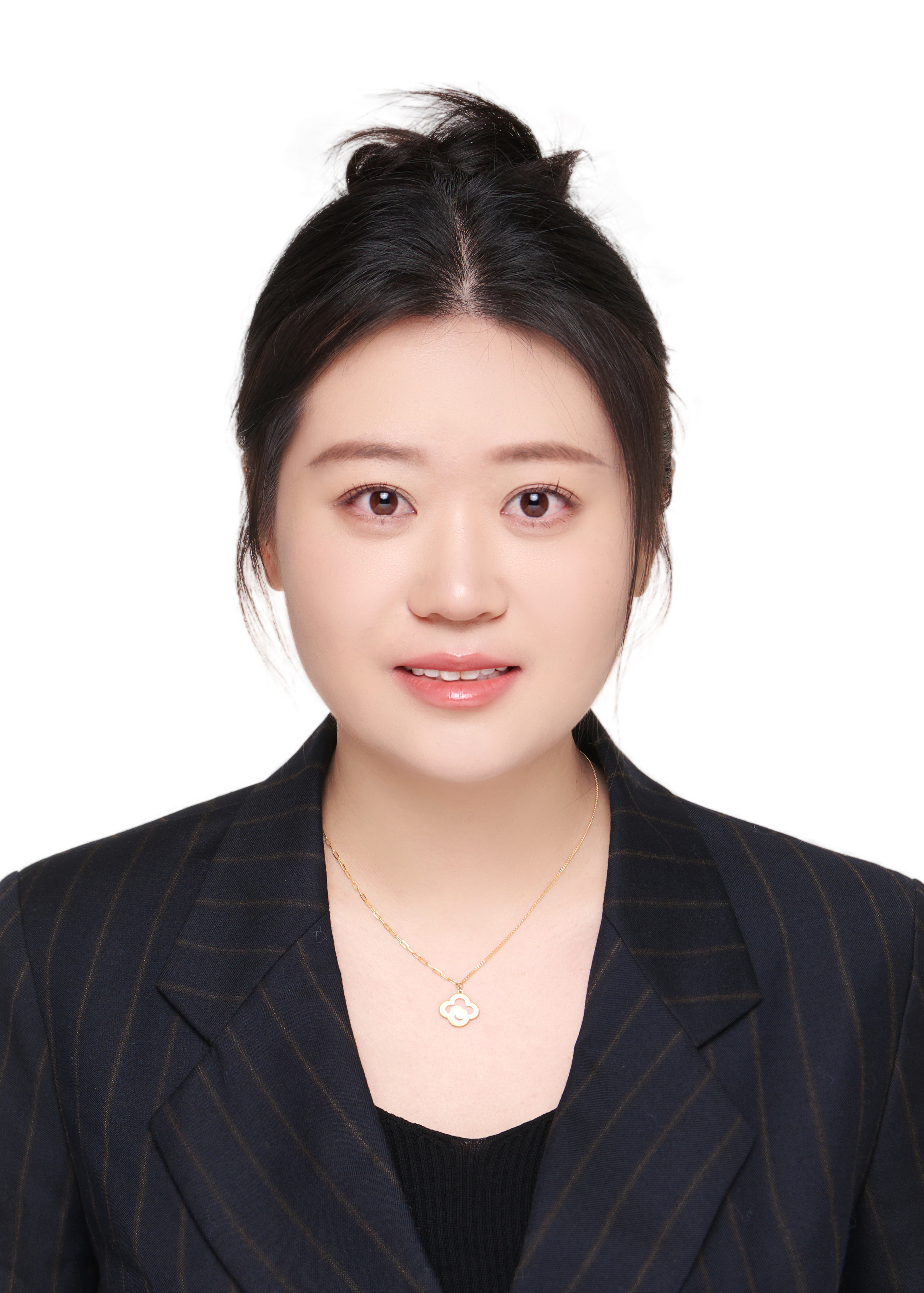}}]{Yujiao Hu}
    (Memeber, IEEE) received her Bachelor and Ph.D. degrees from the Department of Computer Science of Northwestern Polytechnical University, Xi'an, China, in 2016 and 2021 respectively. From Nov. 2018 to March 2020, she was a visiting Ph.D. student in National University of Singapore. Currently, she is a faculty member in Purple Mountain Laboratories. She focuses on deep learning, edge computing, multi-agent cooperation problems and time sensitive networks.    
\end{IEEEbiography}

\begin{IEEEbiography}[{\includegraphics[width=1in,height=1.25in,clip,keepaspectratio]{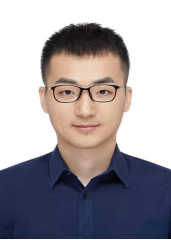}}]{Qingmin Jia} is currently a Researcher in Future Network Research Center of Purple Mountain Laboratories. He received the B.S. degree from Qingdao University of Technology in 2014, and received the Ph.D. degree from Beijing University of Posts and Telecommunications (BUPT) in 2019. His current research interests include edge computing, edge intelligence, IoT and future network architecture. He has served as a Technical Program Committee Member of IEEE GLOBECOM 2021, HotICN 2021. 
\end{IEEEbiography}

\begin{IEEEbiography}[{\includegraphics[width=1in,height=1.25in,clip,keepaspectratio]{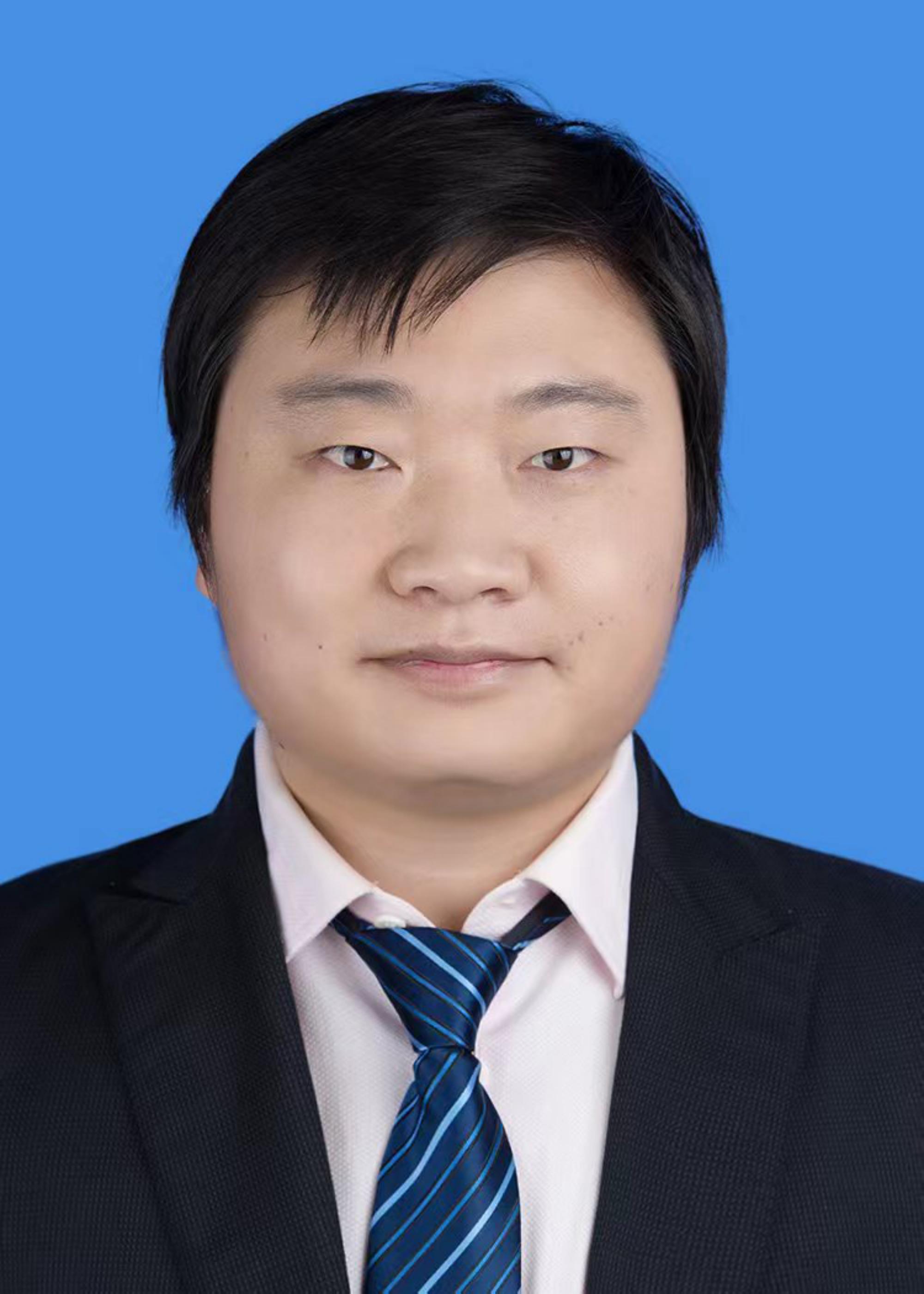}}]{Jinchao Chen} is an associate professor at School of Computer Science in Northwestern Polytechnical University, Xi’an, China. He has received his Ph.D. degree in Computer Science from the same institution in 2016. His interest includes the multi-agent cooperation, real-time system, cloud computing. 
\end{IEEEbiography}

\begin{IEEEbiography}[{\includegraphics[width=1in,height=1.25in,clip,keepaspectratio]{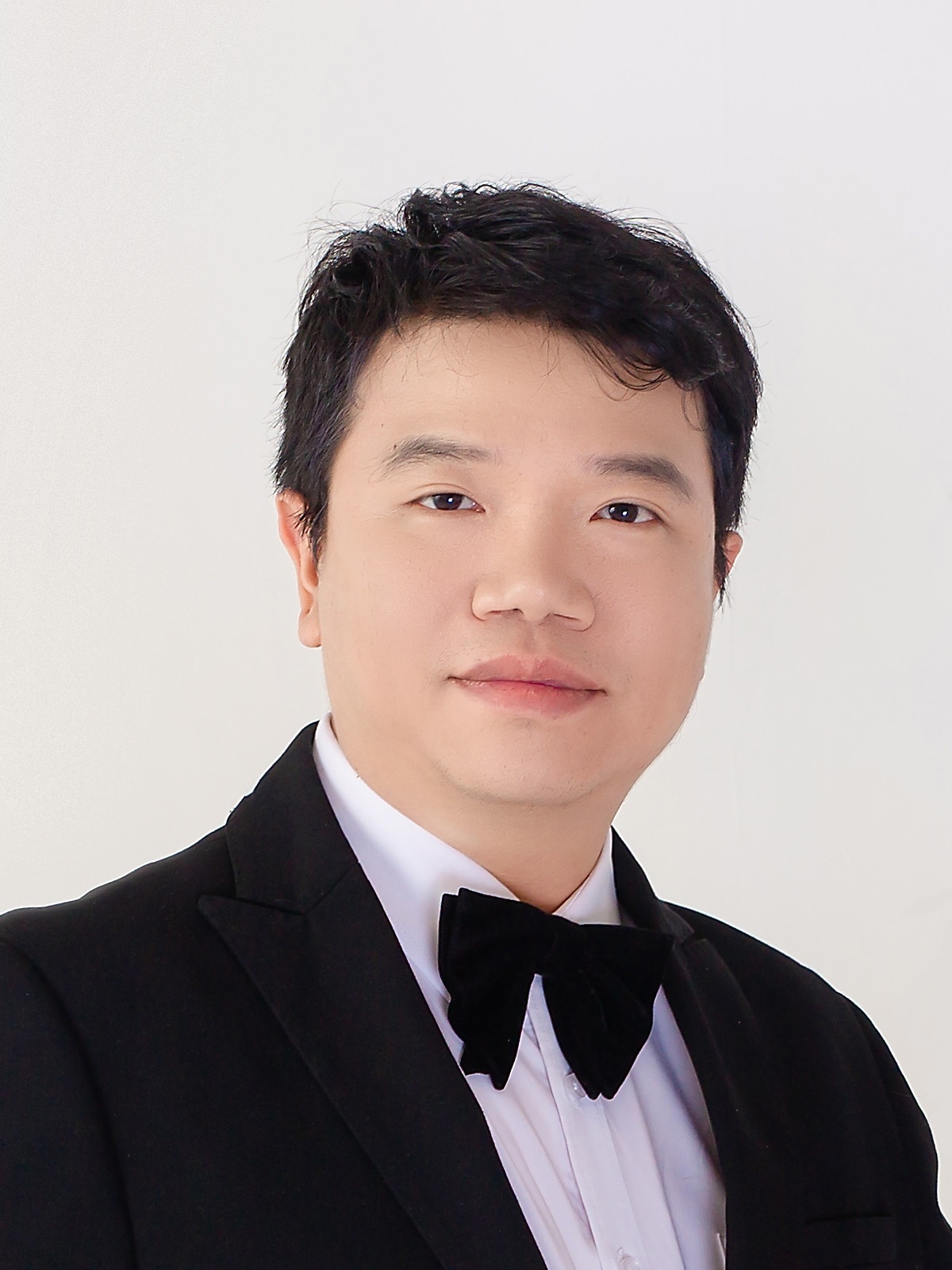}}]{Yuan Yao} is currently an associate professor in the School of Computer Science, Northwestern Polytechnical University (NPU). He received the B.S. M.S., and Ph.D. degrees in computer science from Northwestern Polytechnical University, Xi'an, China, in 2007, 2009 and 2015, respectively. He was a Research Fellow in the Department of Computing at Polytechnic University, Hong Kong from 2016 to 2018. His research interests include real-time and embedded systems, swarm intelligence operating systems and cyber physical system.
\\~\\~\\~\\
\end{IEEEbiography}

\begin{IEEEbiography}[{\includegraphics[width=1in,height=1.25in,clip,keepaspectratio]{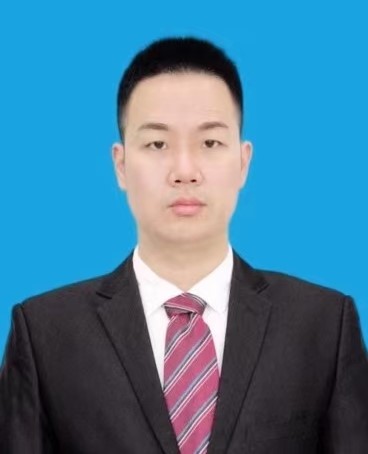}}]{Yan Pan} 
	 is currently a lecture at the Science and Technology on Information Systems Engineering Laboratory, National University of Defense Technology, Changsha, China, since Dec. 2020. Before that, he respectively received the B.S. degree in 2013 and the Ph.D. degree in 2021 from Northwestern Polytechnical University, Xi'an, China. He was a visiting student to University of Maryland, United States, during Jan. 2017 and Nov. 2018. His research interests include Industrial Internet of Things, industrial robots, and edge computing.
\end{IEEEbiography}

\begin{IEEEbiography}[{\includegraphics[width=1in,height=1.25in,clip,keepaspectratio]{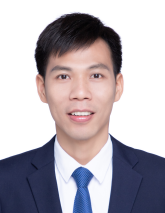}}]{Renchao Xie} received the Ph.D. degree from the School of Information and Communication Engineering, Beijing University of Posts and Telecommunications (BUPT), Beijing, China, in 2012. He is a Professor with BUPT. From July 2012 to September 2014, he worked as a Postdoctoral Fellow with China United Network Communications Group Company. From November 2010 to November 2011, he visited Carleton University, Ottawa, ON, Canada, as a Visiting Scholar. His current research interests include 5G network and edge computing, information-centric networking, and future network architecture. Dr. Xie has served as a Technical Program Committee Member of numerous conferences, including IEEE Globecom, IEEE ICC, EAI Chinacom, and IEEE VTC-Spring.
\\~\\
\end{IEEEbiography}

\begin{IEEEbiography}[{\includegraphics[width=1in,height=1.25in,clip,keepaspectratio]{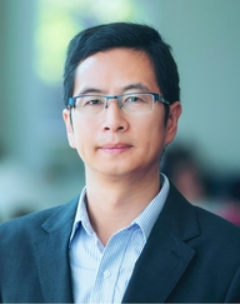}}]{F.Richard Yu}
received the Ph.D. degree in electrical engineering from the University of British Columbia, Vancouver, BC, Canada, in 2003. From 2002 to 2006, he was with Ericsson, Lund, Sweden, and a start-up in California, USA. He joined Carleton University, Ottawa, ON, Canada, in 2007, where he is currently a Professor. His research interests include wireless cyber–physical systems, connected/autonomous vehicles, security, distributed ledger technology, and deep learning. He is a Distinguished Lecturer, the Vice President (Membership), and an Elected Member of the Board of Governors of the IEEE Vehicular Technology Society. He is a Fellow of the IEEE, Canadian Academy of Engineering (CAE), Engineering Institute of Canada (EIC), and Institution of Engineering and Technology (IET).
\end{IEEEbiography}

\end{document}